\documentclass[twocolumn,prb,aps,superscriptaddress,showpacs]{revtex4-1}

\usepackage{color}
\usepackage{times}
\usepackage{latexsym,amsmath,amssymb,bm,euscript}
\usepackage{amsmath}
\usepackage{amssymb}
\usepackage{graphicx}
\usepackage{esint}
\usepackage[unicode=true,pdfusetitle,
 bookmarks=true,bookmarksnumbered=false,bookmarksopen=false,
 breaklinks=false,pdfborder={0 0 1},backref=section,colorlinks=false]
 {hyperref}
\hypersetup{
 colorlinks,citecolor=blue,filecolor=blue,linkcolor=blue,urlcolor=blue}
\usepackage{breakurl}

\makeatletter

\begin{document} 

\title{Electronic and optical properties of InGaAs quantum wells with Mn-delta-doping GaAs barriers} 

\author{Udson C. Mendes}
\email{udsonmendes@gmail.com}
\thanks{Present Address: Laboratoire Pierre Aigrain, \'{E}cole Normale Sup\'{e}rieure, Paris.}
\author{M. A. G. Balanta}
\author{Maria J. S. P. Brasil}
\author{Jos\'{e} A. Brum}
\affiliation{Instituto de F{\'{i}}sica ``Gleb Wataghin'', Unviersidade Estadual de Campinas, UNICAMP, 13083-859 Campinas, S\~{a}o Paulo, Brazil}
\date{\today} 

\begin{abstract}
We present here the electronic structure and optical properties of InGaAs quantum wells with barrier doped with Manganese. We calculated the 
electronic states and optical emission within the envelope function and effective mass approximations using the spin-density functional theory 
in the presence of an external magnetic field. We observe magneto-oscillations of the Landau levels at low-magnetic fields ($B< 5$ T) that are 
dominated by the magnetic interaction between holes spin and Mn spin, while at high magnetic fields the spin-polarization of the hole gas is the dominant effect. Our results also show that a gate voltage alter significantly the magneto-oscillations of the emission energy and may be an 
external control parameter for the magnetic properties of the system. Finally, we discuss the influence of the Landau Levels oscillations in the emission spectra and compare with available experimental.
\end{abstract}

\maketitle 

\section{Introduction}
\noindent 

The research on magnetic semiconductors has attracted much attention
for more than two decades \cite{Munekata1989,Dietl2000,Jungwirth2013a}.
The most investigated material is the (Ga,Mn)As system, where it has
been observed ferromagnetic phase with Curie temperatures ($T_{c}$)
reaching 190 K for samples with Mn concentration of $\sim10\%$.\cite{Olejnik2008,Chen2009}
Others (III,Mn)V materials, such as (In,Mn)As, (Ga,Mn)Sb, (In,Mn)Sb, 
have also shown ferromagnetic phase \cite{dietl-ohno-rmp2014}. In these materials,
Mn acts as both an acceptor and a magnetic impurity and its ferromagnetism is mediated by the 
interaction between holes and Mn spins \cite{Dietl2000,Jungwirth2006}. 
As the magnetic interactions are mediated by charged carriers, the control 
of the magnetic properties can be achieved by electrical and optical means 
\cite{Chiba2008,Nemec2012,Tesarova2013}.

Much of the research efforts have been concentrated on (III,Mn)V bulk or in its heterostructures 
where both Mn and holes are in the same spatial region. These structures allow a high ferromagnetic 
Curie temperature due to the strong interaction between the hole gas and the Mn ions. However, the 
hole gas is strongly scattered by the Mn ions, reducing its mobility and the optical quality. 
To overcome this difficulty, the (Ga,Mn)As layers were grown in the presence of a quantum well, 
as for example in a GaAs-(In,Ga)As-(Ga,Mn)As sequence. In this situation, the hole gas is located 
in the quantum well, separated from the Mn ions which need to be controlled in 
order to maintain a certain level of overlap between the holes and the Mn ions to assure the 
magnetic properties. Recently, such heterostructures have been investigated by means of transport \cite{Wurstbauer2010,Knott2011} 
and optical experiments \cite{Zaitsev2010,Gazoto2011,Korenev2012,Balanta2013,balanta-brasil-jap2014}.
The results suggest that the interaction between holes and Mn is determinant for these systems properties.

Gazoto \textit{et al.} \cite{Gazoto2011} investigated (In,Ga)As QWs with GaAs barriers $\delta$-doped 
with both carbon and Mn in alternate sides of the QW. The samples were $\delta$-doped with Mn in order 
to increase the Mn doping concentration beyond the solubility limit. The presence of the $\delta$-doped 
C layer in the other side of the QW aimed to increase the hole gas concentration and, with that, to 
increase the magnetic effects. They observed that the circularly polarized magneto-photoluminescence 
presents strong oscillations with the magnetic field and they are more pronounced in the samples 
with higher Mn concentration. These oscillations were attributed to the Landau level filling factor.
Magneto-oscillations of the circularly polarized emission were also observed in both two-dimensional 
electron (2DEG) and hole gas (2DHG) \cite{kerridge_brum_ssc1999,Kunc2010,Kehoe2003}. The origin of 
these oscillations is in the many-body effects of both the two-dimensional gas \cite{Uenoyama1989,katayama_ando_ssc1989}
and in the optical recombination process \cite{Hawrylak1997,Takeyama1999,Asano2002}. The oscillations 
of the transition energies observed in Ref. \onlinecite{Gazoto2011} are much stronger than those 
observed in other high quality 2DHGs \cite{Ponomarev1996,Kehoe2003}. These oscillations were correlated 
to the presence of Mn spins in the heterostructure. Several other experiments such as low-magnetic 
field circularly polarized, photoluminescence excitation and time-resolved photoluminescence 
\cite{Zaitsev2010,Korenev2012,Balanta2013,balanta-brasil-jap2014} and transport measurements \cite{Aronzon2010}, 
were performed in similar heterostructures. They all showed significant magnetic interaction 
correlated to the presence of the hole gas.

Here, we present the results of a calculation for the electronic structure and the emission energy of (In,Ga)As 
QWs with GaAs barriers $\delta$-doped with Mn and C. The electronic states of the heterostructure are calculated using 
the spin-density function theory (SDFT) \cite{Hohenberg1964,Kohn1965,Gunnarsson1976,Gupta1982} within the envelope function 
and effective mass approximations \cite{bastard_book}. The Mn-hole spin interaction is described by the Zener kinetic 
exchange theory \cite{Dietl2000,Abolfath2001,Jungwirth2006}. We aim to provide useful information to understand this 
system and to obtain a microscopic interpretation of the observed effects. We compare our results with the experimental 
results from Gazoto \textit{et al.} \cite{Gazoto2011} and suggest an interpretation for the observed magneto-oscillations 
they observe in the emission spectra.

The paper is organized as follows. In Section~\ref{model} we present the model
used to obtain the electronic states of the structure and the emission 
energies. In Sec. \ref{results} we discuss in details the theoretical results 
focusing on the samples used by Gazoto \textit{et al.} \cite{Gazoto2011}. 
Finally, in Sec. \ref{conclusao} we present our concluding remarks. 

\section{Model \label{model}}
\noindent 

\begin{figure}[!htb]
\begin{center}
\includegraphics[width=0.5\textwidth]{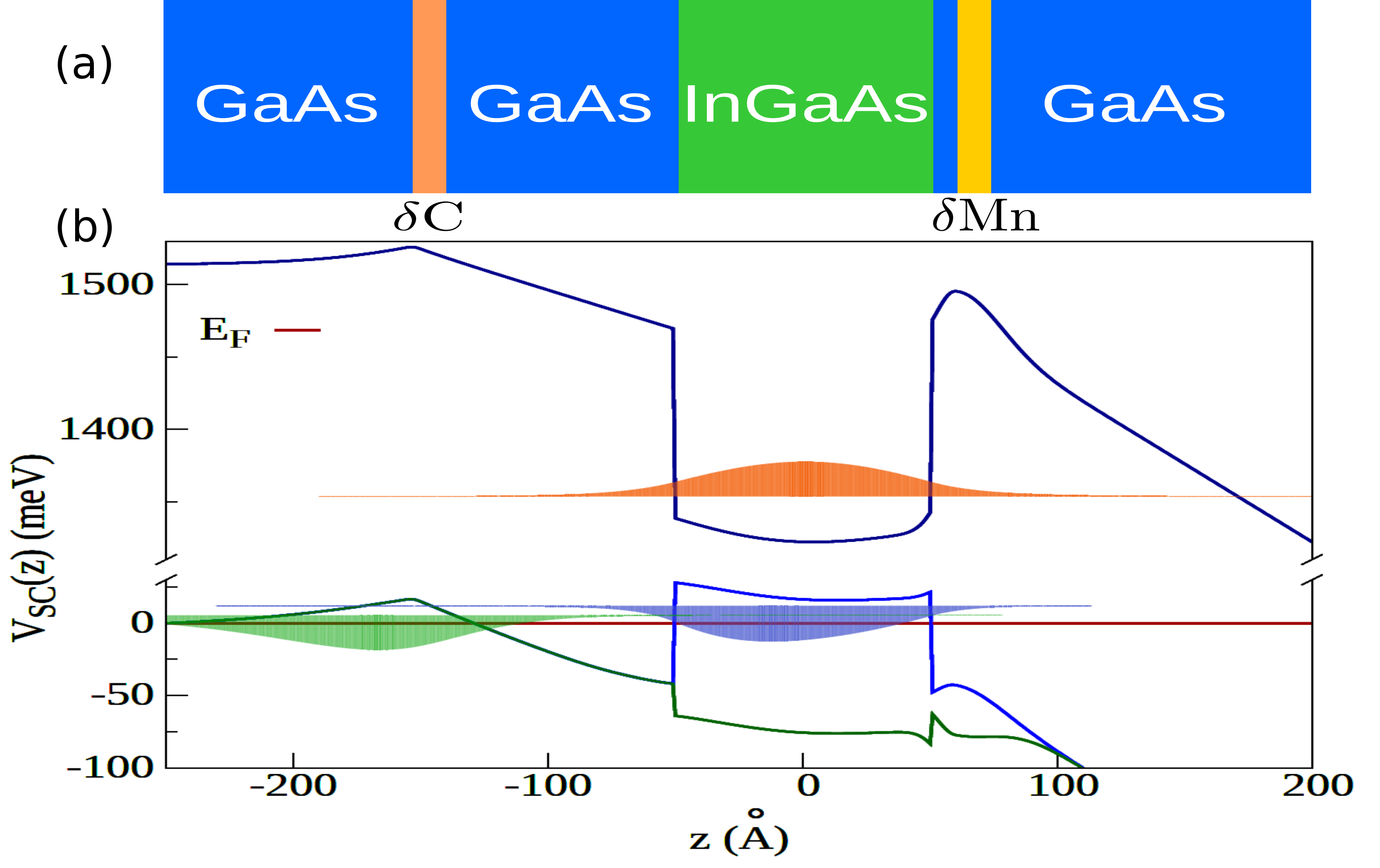}
\caption{(Color online) Schematic representation of the investigated heterostrucutre 
and its self-consistent potential profile and wave-functions. \label{fig1}}
\end{center}
\end{figure}

The model heterostructure investigated is illustrated in Fig.~\ref{fig1}(a) 
and it is based on the system studied by Gazoto {\it et al}  \cite{Gazoto2011}.
It is composed by a 500 nm GaAs buffer layer, a carbon (C) $\delta$-doping
layer, followed by 10 nm GaAs spacer, a 10 nm In$_{0.17}$Ga$_{0.83}$As
layer, a GaAs spacer $L_{s}$, a Mn $\delta$-doping layer with a
concentration $x_{Mn}$ given in percentage of monolayers (MLs), and
finally a 60 nm GaAs top layer. The distance $L_{s}$ between the
QW and the Mn doping layer and the Mn concentration will be considered as free parameters
in our model to study the effects of the Mn interaction. All the others parameters are fixed.
As a consequence of the Mn and C $\delta$-doping and the thermodynamic
equilibrium, the QW presents a 2DHG. We consider the system under
an external magnetic field applied along the growth direction, here
the $z$-direction.

We used the envelope function and effective mass approximations to describe the electronic 
states of the QW \cite{bastard_book}. The alloy is treated in the Virtual Crystal Approximation. 
In general to obtain the valence band (VB) states it is necessary to use the six-band 
Luttinger-Kohn Hamiltonian \cite{Luttinger1955} that describes the heavy-hole ({\it hh}), 
light-hole ({\it lh}) and split-off ({\it so}) bands and the coupling among them as well as the 
spin-orbit effects \cite{Jungwirth2006}. However, in our system, the In$_{0.17}$Ga$_{0.83}$As 
layer is under compressed stress which splits the {\it hh} and the {\it lh} bands by a value of 
the order of 50 meV. The {\it lh} band is actually a type-II or marginally type-I heterostructure, 
depending on the parameters chosen to describe the structure. This results in a near parabolic 
dispersion for the {\it hh} QW ground state for the energies of interest to study the optical emission. 
We therefore simplify our approximation considering a simple parabolic dispersion and tested it against 
a full Luttinger Hamiltonian calculation at zero magnetic field. The comparison is very good for the 
ground state responsible for the optical emission. As we go for the excited states, some discrepancies 
may be present but they do not alter the main results we show here.

The many-body effects of the 2DHG are considered within the SDFT \cite{Barth1972,Gunnarsson1976}. 
This allows us to calculate the ground-state properties of the 2DHG including exchange and correlation 
(XC) effects in the presence of a spin-dependent potential. We employed the Kohn-Sham minimization 
scheme \cite{Kohn1965} to obtain the electronic structure. This procedure maps the many-body problem 
in a set of non-interacting equations, which are solved self-consistently.

We approximate the sample as being homogeneous in the plane. The $z$-direction and the in-plane ($x,y$) 
directions are therefore not coupled. The Hamiltonian can be written as 
\begin{equation}
H^{hh(lh)}=H_{z}^{hh(lh)}+H_{xy}^{hh(lh)},
\end{equation}
where the first term is the $z$-part of the Hamiltonian and the second term is the in-plane Hamiltonian. 
The $xy$-part of the Hamiltonian is responsible for the formation of the Landau Levels (LLs) \cite{Bastard1986}. 
The Hamiltonian in the $z$-direction is 
\begin{align}
H_{z}^{{\it hh(lh)}} & =-\frac{\hbar^{2}}{2m_{{\it hh(lh)}}}\frac{d^{2}}{dz^{2}}+v_{\rm het}^{{\it hh(lh)}}(z)+v_{\rm H}(z)+v_{\rm XC}(z)\nonumber \\
 & +g^{*}\mu_{B}\tau_{z}^{{\it hh(lh)}}B+V_{\it{pd}}^{{\it hh(lh)}}(z).\label{HH_ham}
\end{align}
The first term is the kinetic energy, the second is the heterostructure
potential, the third and fourth terms are the Hartree and XC potentials, respectively.
The fifth is the Zeeman contribution, and lastly, the hole-Mn
(h-Mn) coupling. We will now describe each of the Hamiltonian terms in
details.

\textit{Heterostructure potential} - $v_{\rm het}^{\it hh(lh)}(z)$ is the
structural potential which is built up from the band gap difference and band 
alignment between Ga$_{0.83}$In$_{0.17}$As and GaAs layers plus the strain effects. 
The energy gaps, at low-temperature, are 1.264 eV \cite{Mace1988} and 1.519 
eV \cite{Vurgaftman2001} for Ga$_{0.83}$In$_{0.17}$As and GaAs, respectively. 
The band alignment is type-I. As the QW is strained, its total band offset 
contains both the band gap alignment and strain contributions. We first define 
a band offset without strain \cite{Mace1988}. In this case, we assume a VB (CB) 
offset, $\Delta_{V}$ ($\Delta_{C}$) of 15\% (85\%) of the energy gap difference. \cite{Mace1988}.

Our system is dominated by the GaAs layers and we assume that the
whole structure presents the GaAs lattice parameter. This generates 
a compressive biaxial strain in the Ga$_{0.83}$In$_{0.17}$As layer, 
which alters the QW band offset \cite{Arent1989,Chuang1995}.
If we neglect the {\it so} band, $\Gamma_{7}$, the effect of the
compressive biaxial strain is manifested in an hydrostatic term ($\delta E_h$), which
increases the gap, and a shear deformation ($\delta E_s$), which splits the {\it hh}
and {\it lh} bands \cite{Chuang1995}. The deformation potential for
the hydrostatic term can be split in a contribution to the conduction
band, $\delta E_h^{\rm CB}$ and one to the valence band, $\delta E_h^{\rm VB}$. The final 
result is a QW and a barrier potential for the {\it hh} and {\it lh} bands in the
Ga$_{0.83}$In$_{0.17}$As, respectively. The {\it hh} and {\it lh} potential are given by
\begin{equation}
V_{\rm QW}^{{\it hh/lh}}=\Delta_V-\delta E_h^{\rm VB} \pm \delta E_s ,
\end{equation}
where 
\begin{align*}
\delta E_{h}^{\rm VB} & =a_{\rm VB}(\epsilon_{xx}+\epsilon_{yy}+\epsilon_{zz})\\
\delta E_{s} & =b_{v}(\epsilon_{xx}+\epsilon_{yy}-2\epsilon_{zz}).
\end{align*}
For compressive biaxial strain, the strain components are given by
$\epsilon_{xx}=\epsilon_{yy}=(a_{\text{GaAs}}-a_{\text{GaInAs}})/a_{\text{InGaAs}}<0$,
and $\epsilon_{zz}=-2C_{21}\epsilon_{xx}/C_{11}$, where $a_{\rm GaAs}$
and $a_{\rm GaInAs}$ are the GaAs and Ga$_{0.83}$In$_{0.17}$As
lattice parameters, respectively. $a_{\rm VB(CB)}$ and $b_{v}$ are the
deformation potentials. $C_{11}$ and $C_{21}$ are elastic stiffness
constants \cite{Chuang1995,Vurgaftman2001,Arent1989}.

We can now turn our attention to the insertion of the Mn in the GaAs. The Ga$_{1-x_{\rm Mn}}$Mn$_{x_{\rm Mn}}$As 
layer has a larger lattice parameter than GaAs and follows a similar analysis as for the In$_{0.17}$Ga$_{0.83}$As 
layer regarding the strain effects. One additional difficulty is that the Mn strongly diffuses towards the 
surface in the GaAs as it was shown from secondary ion mass spectroscopy (SIMS) \cite{Poggio2005,Nazmula2003,Wurstbauer2009}. 
Based on the SIMS results, instead of considering a Mn $\delta$-doping layer, we
assumed that the total Mn is distributed over many GaAs layers, forming
a Ga$_{1-x_{\rm Mn}}$Mn$_{x_{\rm Mn}}$As alloy. To take this effect into
account, we assume that the system is homogeneous in the $x$-$y$
directions and only variations of the Mn concentration in the $z$-direction
are considered. We construct a distribution function that takes in
account the Mn diffusion in the GaAs \cite{Poggio2005,Nazmula2003,Wurstbauer2009}. 
The SIMS results show that the Mn diffuses following an approximate gaussian-like function in both directions 
of the heterostructure. However, it diffuses more strongly in the
direction of the surface than towards the QW. We consider
therefore a double-gaussian distribution function, as defined below
\begin{equation}
f(z-L_{s})=f_{0}exp\{-[(z-L_{s})/\Delta(z)]^{2}\},\label{difusao}
\end{equation}
where $L_{s}$ is the gaussian center, that is, the nominal $\delta$-Mn
doping position. $\Delta(z)$ is the average width of the gaussian
that describes the Mn diffusion 
\begin{equation}
\Delta(z)=\begin{cases}
d\:\:\text{if}\:\: z<L_{s},\\
D\:\:\text{if}\:\: z\geq L_{s},
\end{cases}
\end{equation} 
and $f_{0}$ is the normalization constant. We considered $D=2$ nm and $d=1$ nm, which are 
compatible with the SIMS results \cite{Poggio2005,Nazmula2003,Wurstbauer2009}. 
The Ga$_{1-x_{\rm Mn}}$Mn$_{x_{\rm Mn}}$As layers also shows a different gap than GaAs, 
and an intrinsic band offset should be present. The main contribution for the 
Ga$_{1-x_{\rm Mn}}$Mn$_{x_{\rm Mn}}$As band offset, however, has origin in the 
$sp$-$d$ interaction, which is discussed below. We therefore will neglect the intrinsic 
band offset.

Finally, we consider the $\delta$-C layer. C act as an acceptor and its growth 
is well controlled and does not significantly diffuse. We assume an homogeneously 
distributed 5 \AA{} doped region in a continuous approximation. The heterostructure 
potential finally can be written as
\begin{align*}
v_{\rm het}^{hh/lh}(z) & =V_{\rm QW}^{\it hh/lh}\Theta(z^{2}-L_{\rm QW}^{2}/4)\\
& -(\delta E_{h_{\rm Mn}}^{\rm VB}\mp \delta E_{sh}^{\rm Mn})a_{\rm GaMnAs}f(z-L_{s})\\
\end{align*}
with $\Theta(x)$ being the Heaviside function, and $L_{\rm QW}$ the
QW width, that is, the In$_{0.17}$Ga$_{0.83}$As layer.

\textit{Hartree potential} - The third term of $H_{z}$ is the Hartree
potential, which is obtained by solving Poisson's equation 
\begin{align}
\frac{d^{2}v_{\rm H}(z)}{dz^{2}} & =-\frac{e^{2}}{4\pi\epsilon}\left[p(z)-N_{c}\Theta(z-z_{c}^{f})\Theta(z_{c}^{0}-z)\right. \label{poisson}\\
 & \left.-p_{Mn}f(z-L_{s})\right], \nonumber 
\end{align}
where $\epsilon$ is the GaAs dielectric constant. $p(z)$ is the
total hole density, which at zero temperature ($T=0$ K) is given
by 
\begin{equation}
p(z)=\begin{cases}
{\displaystyle \sum_{\substack{b,i\\
\tau_{z}
}
}}\frac{m_{p}^{b}}{2\pi\hbar^{2}}|\psi_{i,\tau_{z}}^{b}(z)|^{2}\Delta_{i,\tau_{z}}^{b}\Theta(\Delta_{i,\tau_{z}}^{b})\:\:\text{if}\:\: B=0\text{T},\\
{\displaystyle \sum_{\substack{b,i\\
n,\tau_{z}
}
}\frac{m_{p}^{b}}{2\pi\hbar^{2}}}|\psi_{i,\tau_{z}}^{b}(z)|^{2}\int_{-\infty}^{E_{F}}g_{i,\tau_{z}}^{b,n}(\varepsilon)d\varepsilon\:\:\text{if}\:\: B\neq0\text{T}.
\end{cases}\label{density}
\end{equation}
where
\begin{equation}
g_{i,\tau_{z}}^{b,n}(\varepsilon)=\frac{eB}{2\pi\hbar}\frac{1}{\sqrt{2\pi}\Gamma}\text{exp}\left[-\frac{(\varepsilon-E_{i,\tau_{z}}^{b,n})^{2}}{2\Gamma^{2}}\right],
\end{equation}
The first (second) line describes the hole density in absence (presence) of an external magnetic field. $\psi_{i,\tau_{z}}^{b}(z)$ are the
VB envelope functions, and $\Delta_{i,\tau_{z}}^{b}=E_{F}-E_{i,\tau_{z}}^{b}$, where $E_{F}$ is the Fermi level, and $E_{i,\tau_{z}}^{b}$ 
the eigenvalues of $H_z^b (b={\it hh,lh}$). $E_{i,\tau_{z}}^{b,n}=E_{i,\tau_{z}}^{b}+E_{n}^{b}$ is the total subband energy of the $b$-hole with 
spin $\tau_{z}$ in the $n$-th LL of the $i$-th subband. $\Gamma=\Gamma_{0}\sqrt{B}$ is the LL broadening related to the 2DHG mobility \cite{bastard_book,Ando1974}. In our calculations, we considered $\Gamma_{0}$ as a parameter.

In the absence of a gate voltage, we assume that the Fermi level, $E_{F}$,
is pinned at the surface states, that is, in the middle of the gap
at the surface, $E_{F}=E_{g}^{GaAs}/2$ \cite{dietl-ohno-rmp2014,sampaio_alves_jap1997}.
We will also consider the case of an applied gate voltage, $V_{g}$,
which allows us to change the Fermi level position. In this case we have 
$E_{F}=E_{g}^{GaAs}/2-V_{g}$.

\textit{Exchange-correlation potential} - The forth term of Eq.
(\ref{HH_ham}) is the XC potential. Here we use the Vosko, Wilk,
Nusair (VWN) parametrization \cite{Vosko1980,perdew_wang_prb1992} for the local-spin-density
approximation. $v_{\rm XC}(z)$ depends on both the hole density $p(z)=p_{\uparrow}(z)+p_{\downarrow}(z)$
and the hole gas magnetization $\xi(z)=p_{\uparrow}(z)-p_{\downarrow}(z)$.
Again, we consider that the density in the plane is homogeneous,
and hence, the hole density depends only on the $z$-coordinate.

\textit{Zeeman potential} - The fifth term in $H_{z}$ is the Zeeman
interaction between hole spins and the external magnetic field. $g^{*}$ is 
the hole effective g-factor and $\mu_{B}$ the Bohr magneton. The {\it hh}
and {\it lh} spins are $\tau_{z}^{hh}=\pm3/2$ and $\tau_{z}^{lh}=\pm1/2$,
respectively. 

\textit{$p$-$d$ potential} - The last term of the Hamiltonian is
the $p$-$d$ interaction between holes and Mn spins. This term has
its origin in the interaction between VB states with the $d$-orbitals
of the Mn impurity \cite{larson_carlson_prb1988,Kacman2001}. The presence
of the hole gas is described via Zener kinetic-exchange model \cite{Dietl2000,Abolfath2001,Jungwirth2006}.

We assumed the magnetization vector aligned along the $z$-direction. 
This is a valid approximation since even a small magnetic field aligns the 
magnetization in its direction \cite{Dietl2001} against the easy-axis \cite{Abolfath2001,Dietl2001}.

The final expression is written as 
\begin{equation}
V_{pd}^{hh(lh)}(z)=-\frac{1}{3}N_{0}\beta x_{eff}M\tau_{z}^{hh(lh)}\mathcal{B}_{M}(y)a_{Mn}f(z-L_{s}),
\end{equation}
where $N_{0}\beta$ is the $p$-$d$ exchange constant of the spin
interaction between Mn's and holes. $x_{eff}$ is the effective concentration of Mn spins (see below).
$M=5/2$ is the Mn spin. $\mathcal{B}_{M}(y)$ is the Brillouin function.\cite{Ashcroft}
Its argument is given by 
\begin{equation}
y=\frac{g_{Mn}\mu_{B}MB}{k_{B}T}+\frac{J_{pd}M}{2k_{B}T}\int\xi(z)f(z-L_{s})dz,\label{Brillouin}
\end{equation}
the first term is due to the interaction of Mn spin with the external
magnetic field, where $g_{Mn}$ is the Mn g-factor, $k_{B}$ is the
Boltzmann constant, and $T$ is the temperature. The second term is
the antiferromagnetic interaction between holes spins with Mn spins, 
which is responsible for the ferromagnetic interaction
of Mn's spins \cite{Dietl2000,Abolfath2001,Jungwirth2006}. $J_{pd}=\beta/N_{0}$
is the {\it p-d} exchange constant. $N_{0}$ is the cation concentration.

This antiferromagnetic interaction depends on the 2DHG magnetization
and the overlap between hole and Mn ions. If Mn ions and holes are 
equally homogeneously distribute, the distribution function $a_{Mn}f(z-L_{s})$ is 
replaced by a unitary constant and we recover the well-known results from 
(Ga,Mn)As bulk \cite{Jungwirth2006}. Here, they are non-homogeneous and the second 
term in the function $y$ depends strongly on the structural parameters, 
namely, the Mn position in the GaAs and the holes states.

Equations \ref{HH_ham}, \ref{poisson}, and \ref{density} are solved self-consistently. 
The Schr{\"{o}}dinger equation is solved via split-operator method \cite{Degani2010}. 

\textit{Conduction band states} - The CB states are calculated within the same approximations. The in-plane Hamiltonian 
gives the LLs for the CB. The electron Hamiltonian in the $z$-direction is 
\begin{align}\label{Hamil_elec}
H_{z}^{e} & =-\frac{\hbar^{2}}{2m_{e}}\frac{d^{2}}{dz^{2}}+v_{\rm het}^{e}(z)-v_{\rm H}(z)+g_{e}\mu_{B}\sigma_{z}B\\
 & +V_{sd}^{e}(z)+v_{\rm C}(z)\nonumber 
\end{align}
where $m_{e}$ is the electron effective mass in the CB. The second
term is the CB heterostructure potential, which can be written as
\begin{align}
v_{\rm het}^{e}(z) & =(\Delta_{\rm C}-\delta E_{h}^{\rm CB})\Theta(z-L_{\rm Qw}/2)\Theta(L_{ \rm Qw}/2-z)\nonumber \\
 & -\delta E_{h_{\rm Mn}}^{\rm CB}a_{\rm GaMnAs}f(z-L_{s}).
\end{align}
The strain contribution in the CB offset is limited to the hydrostatic term
($\delta E_{hy}^{CB}$). The third term is the Hartree
potential defined in Eq. \ref{poisson}. The fourth term is the Zeeman
interaction, where $g_{e}$ is the electron g-factor and $\sigma_{z}=\pm1/2$
is the electron spin. In the fifth term we have the $s$-$d$ interaction
between the electron's and Mn's spins, which is written as 
\begin{equation}
V_{sd}^{e}(z)=N_{0}\alpha x_{eff}M\sigma_{z}\mathcal{B}_{M}(y)a_{Mn}f(z-L_{s}),
\end{equation}
where $N_{0}\alpha$ is the $s$-$d$ exchange constant between electron
and Mn spin. The last term is the effect of the correlation potential 
on the conduction band due to the presence of the hole gas. Since
electrons and holes are treated as different particles, there is no
exchange contribution for the CB. This potential was parametrized
for the case of a spin-unpolarized 2DHG \cite{Bauer1986,Bobbert1997}
and gives an important contribution for the band-gap renormalization
observed in the optical spectrum of modulated-doped QWs \cite{Bauer1986}.
From our knowledge there is no parametrization for the spin-dependent 
electron-hole correlation energy. We will neglect this term in our 
calculations. Within our description the electron-hole correlation 
potential should not present a dependence with the spin-polarization 
of the hole gas, but only on the total hole gas density. Therefore, 
the main effect of this contribution will depend on the total charge
transfer between the hole reservoirs of the heterostructure which plays 
a minor effect in most of our results (see next Section).

\textit{Optical transitions} - The transition energy is calculated as the energy difference between
the electron and hole eigenstates. Our focus is in the circularly
polarized emission. The right circularly polarized ($\sigma_{+}$)
light is given by the recombination of a spin-down electron with a
spin-up {\it hh}, while the left circularly polarized ($\sigma_{-}$)
light is the recombination of a spin-up electron with a spin-down
{\it hh}. The recombination energies are given by 
\begin{align*}
E_{Tot}^{\sigma_{+}(\sigma_{-})} & =E_{j,\downarrow(\uparrow)}^{e}+E_{m}^{e}+E_{i,\uparrow(\downarrow)}^{hh}+E_{n}^{hh},
\end{align*}
The emission is allowed only if the electron and hole states
are in the same LL ($n=m$). To enhance the magnetic field
effects on the transition energies, we subtract the transition energy
at zero magnetic field from $E_{T}^{\sigma_{+}(\sigma_{-})}(B)$,
redefining the transition energy as 
\begin{equation}
E_{T}^{\sigma_{+}(\sigma_{-})}(B)=E_{Tot}^{\sigma_{+}(\sigma_{-})}(B)-E_{Tot}^{\sigma_{+}(\sigma_{-})}(0).
\end{equation}
We define the non-linear energy shift as 
\begin{equation} \label{nonlinear}
\Delta E_{T}^{\sigma_{+}(\sigma_{-})}(B)=E_{T}^{\sigma_{+}(\sigma_{-})}(B)-E_{z}^{\sigma_{+}(\sigma_{-})}(B)-E_{L}(B)
\end{equation}
where we subtracted all the linear terms in $B$ in order to magnify the non-lienar effects in the transition energy.

\textit{Parameters} - The holes in the heterostructure are provided
by both the C and the Mn doping. We will consider the effective Mn concentrations parameters 
that better agree with those from [\onlinecite{Gazoto2011}]. 
The C concentration $N_{c}$ is fixed for all systems we investigate here. 
We consider $N_{c}=13.35\times10^{18}$ cm$^{-3}$ which is the value 
obtained by fitting the measured hole concentration in the QW for 
a sample without Mn and comparing with our calculations \cite{Gazoto2011}. 
The nominal concentration of Mn, $x_{Mn}$, is known from the growth process. 
However, it does not provide the real hole density, since Mn can be 
either a substitutional impurity or an interstitial one \cite{Jungwirth2006}. 
In the first case Mn replaces Ga, and provides one hole to the system, while at
the interstitial position it is a double donor, and gives two electrons.
Therefore, there is a self-compensation of holes by the electrons,
and the total density of holes provided by the Mn is given by $p_{Mn}=x_{S}-2x_{I}$,
where $x_{S}$ and $x_{I}$ are the concentration of substitutional
and interstitial Mn, respectively \cite{Jungwirth2006}. Furthermore,
because of the attractive Coulomb interaction, the interstitial Mn
ions tends to be near to the substitutional ones presenting an antiferromagnetic
coupling, which reduces the net Mn spins \cite{Jungwirth2006}. The
effective Mn spin concentration is given by $x_{eff}=x_{S}-x_{I}$ \cite{Jungwirth2006,dietl-ohno-rmp2014}.
In our model, we only describe the uncompensated substitutional Mn,
$p_{Mn}$, with the effective spin concentration, $x_{eff}$. We do
not have direct access to $x_{S}$ and $x_{I}$. These values are strongly 
dependent on sample growing conditions. We extract $p_{Mn}$ from our 
calculations by fitting the theoretical value $p_{QW}$ with its experimental 
value, which was estimated from Shubnikov-de-Hass and Stoke shift
measurements. This allows to determine $p_{Mn}$ for each sample which is 
used as fixed parameters for the remaining calculations. The other parameters 
are described in Table \ref{table1}.
\begin{table}[!h]
\begin{center}
\caption[InAs and GaAs parameters]{Parameters used in the self-consistent calculation. 
The In$_{0.17}$Ga$_{0.83}$As strain parameters are linear interpolation between GaAs and InAs parameters. 
The parameters were extracted from the refs. \onlinecite{Vurgaftman2001,wimbauer_brugger_prb1994,
okabayashi_tanaka_prb1998,Jungwirth2006,kotlyar_forchel_prb2001,Mace1988,Arent1989}. \label{table1}}
\begin{tabular}{ccc}
\hline 
Parameters  & GaAs  & InAs \tabularnewline
\hline 
\hline 
$\gamma_{1}$\cite{Vurgaftman2001}  & 6.98  & 20.0 \tabularnewline
$\gamma_{2}$\cite{Vurgaftman2001}  & 2.06  & 8.5 \tabularnewline
$C_{11}$ (10$^{10}$ Pa)\cite{Vurgaftman2001}  & 12.21  & 8.329 \tabularnewline
$C_{21}$ (10$^{10}$ Pa)\cite{Vurgaftman2001}  & 5.66  & 4.526 \tabularnewline
$a_{L}$ (\AA{})\cite{Vurgaftman2001}  & 5.65325  & 6.0583 \tabularnewline
$a_{vb}$ (eV)\cite{Vurgaftman2001}  & -7.17  & -5.08 \tabularnewline
$a_{cb}$ (eV)\cite{Vurgaftman2001}  & -1.16  & -1.0 \tabularnewline
$b_{v}$ (eV)\cite{Vurgaftman2001}  & -2.0  & -1.8 \tabularnewline
\hline 
\hline 
$m_{p}^{hh}$\cite{wimbauer_brugger_prb1994} & 0.11 \tabularnewline
$g_{e}$\cite{wimbauer_brugger_prb1994}  & -2.9 \tabularnewline
$g^{*}$\cite{wimbauer_brugger_prb1994}  & -2.3 \tabularnewline

$E_{g}(x_{In})$ (eV)\cite{Mace1988}  & 1.519 - 1.583$x_{In}$ + 475$x_{In}^{2}$  & \tabularnewline
$\Delta_{VB}$ (eV)\cite{Arent1989}  & 0.15$E_{g}(x_{In})$  & \tabularnewline
$N_{0}\alpha$ (eV)  & 0.2  & \tabularnewline
$N_{0}\beta$ (eV)\cite{okabayashi_tanaka_prb1998,Jungwirth2006}  & 1.2  & \tabularnewline
$J_{pd}$ (meV nm$^{3}$)\cite{Jungwirth2006}  & 54  & \tabularnewline
$x_{Mn}$ & 0.4  & \tabularnewline
$x_{eff}$ & 0.13  & \tabularnewline
$p_{Mn}$ ($10^{11}$ cm$^{-2}$) &  9.92  & \tabularnewline
$p_{QW}^{1st}$ ($10^{11}$ cm$^{-2}$) &  5.2  & \tabularnewline
\hline 
\end{tabular}
\end{center}
\end{table}
\section{Results \label{results}}
\noindent 

We consider a sample with $x_{Mn}=0.4$ ML and $L_{s}=1$ nm  as our case study. The LLs broadening ($\Gamma_{0}$) is fixed for all 
investigated heterostructures independently of $x_{Mn}$. $\Gamma_{0}$ is related to the 2DHG mobility, 
which is approximately $\sim2\times10^{3}$ cm$^{2}$/Vs at 77 K in the QW for the samples investigated 
in Ref. \onlinecite{Gazoto2011}. This implies in $\Gamma_{0}\approx1.8$ meV B$^{-1/2}$. The 
experimental photoluminescence is performed at 2 K. At lower temperatures the mobility increases 
and this leads to smaller values for the LLs broadening. In our calculation we considered 
$\Gamma_{0}=0.25$ meV B$^{-1/2}$. Fig. \ref{fig1}(b) illustrates the self-consistent profile potential 
for the {\it hh} and {\it lh} VB. The wave-functions in the VB represent the occupied {\it hh}-subbands. 
We note that there is a 2DHG in the carbon doped layer. In the Mn layer there is no occupied state. 
Therefore, the only hole gas interacting with the Mn ions has its origin in the QW states. 
We assume that the Fermi level is pinned at the surface states which creates a high electric 
field between the surface and the QW as it can be observed in Fig. \ref{fig1}(b). As a consequence 
of this strong potential anisotropy and the different doping at the barriers, the QW CB ground state 
(not shown here) is pushed towards the Mn doped barrier. These features are common for most of the
case discussed here.

\begin{figure}[!htb]
\begin{center}
\includegraphics[width=0.5\textwidth]{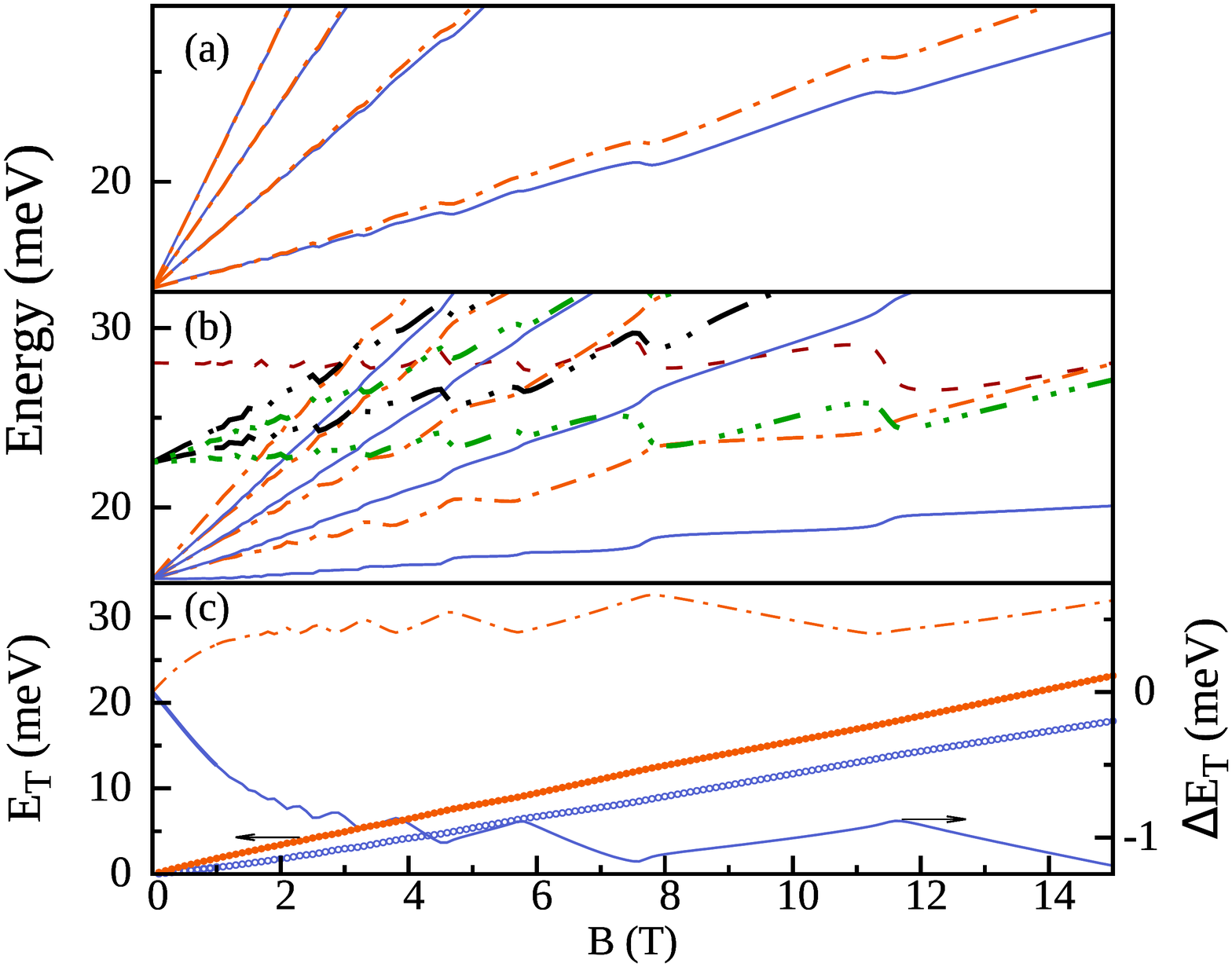} 
\caption{(Color online) (a) Electron Landau levels fan diagram. The solid and dash-dot lines represents the $e_{m,\uparrow}^{1}$ and $e_{m,\downarrow}^{1}$ LLs, respectively.(b) Heavy-hole Landau Level as a function of the magnetic field. Solid, dash-dot, dash-three-dots, dash two-dots, and dashline represents $h_{n,\uparrow}^{1}$, $h_{n,\downarrow,n}^{1}$, $h_{n,\uparrow,n}^{2}$, $h_{n,\downarrow}^{2}$ and $E_{\rm F}$, respectively. 
(c) Transition energy (left $y$-axis) and non-linear energy shift (right $y$-axis) as a function of the magnetic field.  
Solid and dashdot line are the transition energies for $\sigma_{+}$ and $\sigma_{-}$ polarization. The non-linear energies $\Delta E_{\sigma_{+}}$ and $\Delta E_{\sigma_{-}}$ are represented by open and closed circles, respectively.} \label{fig2}
\end{center}
\end{figure}

\begin{figure}[!htb]
\begin{center}
\includegraphics[trim= 0 35 0 60,clip,width=0.5\textwidth]{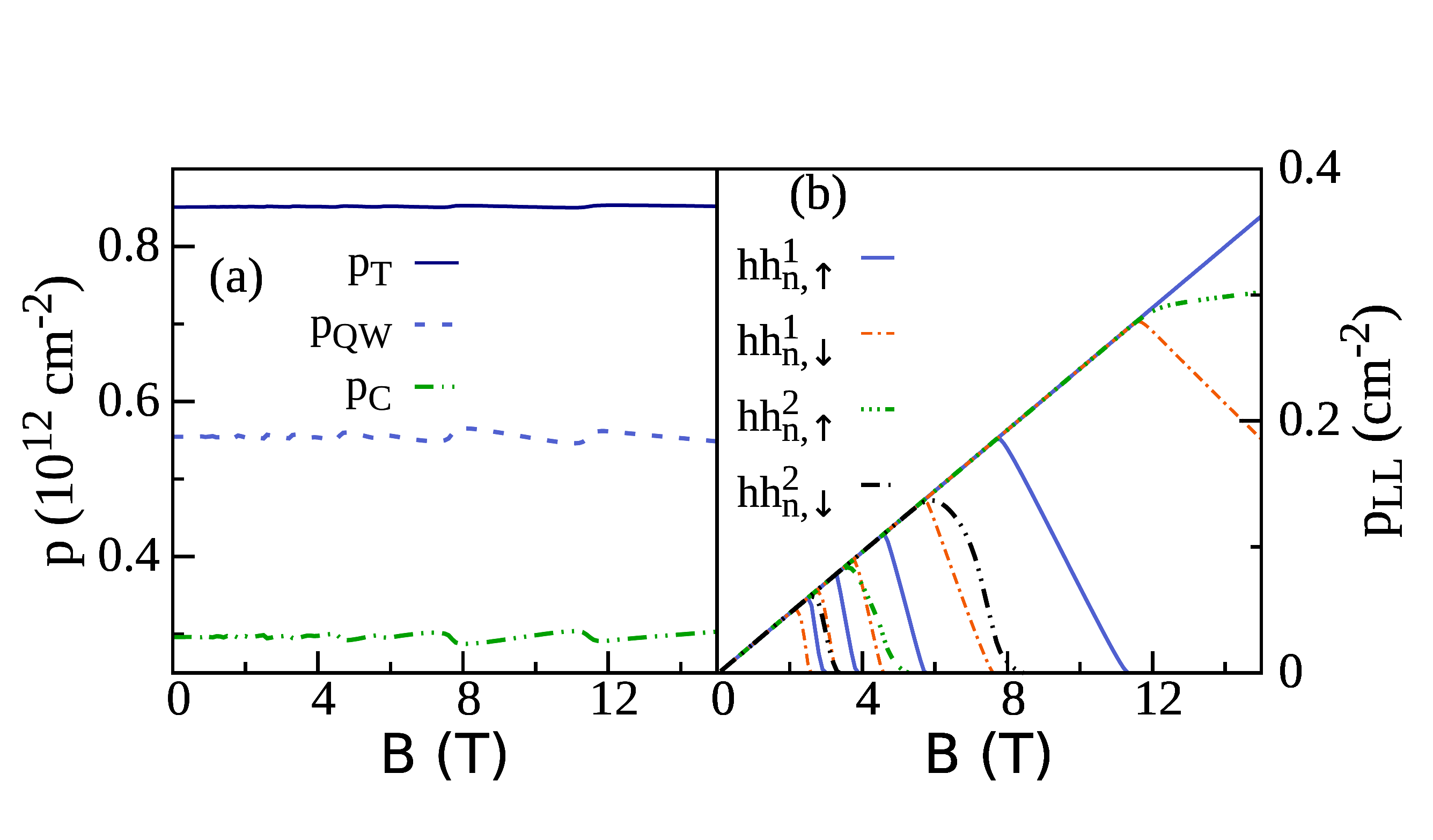} 
\caption{(Color online) (a) Total ($p_{T}$), quantum well ($p_{\text{QW}}$), and carbon ($p_{\text{c}}$) 
two-dimensional hole concentration, and (b) Landau level two-dimensional density as a function of the magnetic field.} \label{fig3}
\end{center}
\end{figure}

First we examine the evolution of the CB LLs as a function of the magnetic field. In Fig. \ref{fig2}(a) 
we plot the LLs associated to the CB QW ground state, $e_{m,\sigma}^{1}$. At low-magnetic fields, more precisely 
at 1.5 T, there is a crossing of the \textit{n=0} CB LLs. This crossing originates from the interplay between 
the Zeeman energy and the \textit{s-d} interaction. It is, however, a too small effect to be observed 
experimentally. At high-magnetic fields, $B>5$ T, we observe oscillations in the CB LLs, which are associated 
to the crossing of the QW hole LLs with the Fermi energy. These oscillations did not change the CB LLs 
dependence with the magnetic field and they show the same behavior for both spins. This is expected because the 
electron spin does not interact with the hole spin via exchange interaction, and also the {\it s-d} exchange 
coupling cause only a rigid energy shift at large magnetic field. Therefore, the observed oscillations have their 
origin in the charge transfer between the C layer and the QW, as it will be discussed bellow. We remind that 
the electron-hole correlation was not taken into account in our model. However, we do not expect a qualitative 
influence from this term on the spin-properties of the QW hole gas. It would certainly lower the energy levels 
and enhance the oscillations since correlation potential should be dependent on the total hole density. Its effect 
is associated to the oscillations in the total hole concentration. As we observed, this is a minor effect [see Fig. 
\ref{fig3}(a)]. The energy transitions calculated within this approximation should give a reliable behavior with the 
magnetic field.

Next we examine the VB electronic structure. Fig. \ref{fig2}(b) shows the LLs fan diagram. In order of increasing energy we have at zero magnetic 
field $hh_{n,\tau_{z}}^{1}$ and $hh_{n,\tau_{z}}^{2}$ subbands, where the upper index, $n$ and $\tau_{z}$ refers to the 
subband order, LL and {\it hh}-spin, respectively. The first and second subbands are located in the QW and 
in the carbon layer, respectively, as depicted in Fig.~\ref{fig1}(b). There is also a marginally occupied light-hole 
subband in the carbon layer, which is not shown in Fig.~\ref{fig1}(b) and does not play any significant role in the 
subsequent analysis. We observe a low- and high-magnetic field regimes for the Landau Levels magnetic-field dependence.

We first consider the low-magnetic field regime, {\it i.e.}, $B < 5$ T. Let us concentrate on $hh_{0,\uparrow}^{1}$ and 
$hh_{0,\downarrow}^{1}$ states. They are the more important states for the emission spectra. In particular, for 
$B \leq 2$ T, $hh_{0,\uparrow}^{1}$ state shows an almost flat dependence with the magnetic field while $hh_{0,\downarrow}^{1}$ 
state shows a stronger one. The origin of this behavior is three fold: the Zeeman effect, {\it p-d} interaction, and the 
spin-polarization of the hole gas, which is enhanced by the hole-hole exchange energy. For low magnetic fields the spin-up 
and spin-down hole densities are nearly the same, and hence, the 2DHG spin-polarization does not influence its magnetic 
field dependence (see discussion below for high-magnetic fields, where this contribution is important). Also, in this regime, 
the Zeeman energy is small and its effect is to split the {\it hh}-spin states by the same amount of energy. Therefore, the 
hole-Mn spin interaction is the dominant effect at low magnetic fields. As the magnetic field increases, the Mn spins are 
aligned and the Brillouin function saturates. The net result is a sizable spin-splitting originated by the \textit{p-d} coupling 
that remains at higher magnetic fields. The flat behavior of $hh_{0,\uparrow}^{1}$ originates from the competition between 
\textit{p-d} interaction and diamagnetic shift of the LLs, which tends to cancel out, while for the $hh_{0,\downarrow}^{1}$ 
they add up. This was verified by turning off {\it p-d} interaction (not shown here) in our model.

We turn our attention now to the high-magnetic field regime, {\it i.e.}, $B>5$ T [see Fig. \ref{fig2}(b)]. 
We observe that all LLs and the Fermi level oscillate as a function of the magnetic field. These oscillations 
have origin in both the crossing of the LLs with $E_{F}$, and the charge transfer between the hole gas in the 
QW and the carbon layer. However, we observe that $hh_{0,\uparrow}^{1}$ and $hh_{0,\downarrow}^{1}$ oscillate 
in the same or opposite directions, depending on which spin hole level is crossing $E_{F}$. On the other way, 
the $hh_{0,\uparrow}^{2}$ and $hh_{0,\downarrow}^{2}$ levels roughly following the $E_{F}$ oscillations. 
Fig. \ref{fig3}(a) shows the QW and the C 2DHG concentration 
as a function of the magnetic field. It gives us a measure of the charge 
transfer between the QW and the C layer as a function of the LL filling factor. 
Fig. \ref{fig3}(b) shows the LLs occupation as a function of the magnetic field. Most 
of the LLs show a similar behavior. Their hole concentration increases linearly with 
the magnetic field, as it is expected from the LL degeneracy. As one LL crosses the 
Fermi level, it starts to be depopulated. If there was no broadening, this should 
be an abrupt decrease. In our case, the broadening makes the depopulation 
of the LL to last a finite range of magnetic field, but with a nearly linear 
decrease. This behavior is consistent for all LLs associated to QW subbands. 
The C layer LLs show a different behavior. As they start to be depopulated, 
they do not follow a linear behavior. Actually, this behavior is associated
with the C layer LLs states that roughly follow in energy the Fermi level,
as if they were partially pinned on it.

Taking that in consideration let us look now in more detail in the QW LLs oscillations and their correlation with 
the LL filling factor. Here again we focus on the analyses of the lowest energy QW LLs, {\it i.e.}, $hh_{0,\uparrow}^{1}$ 
and $hh_{0,\downarrow}^{1}$ since they are responsible for the optical emission. At $B \sim$ 5 T the states at the 
QW that are fully occupied are $hh_{i,\tau_{z}}^{1}$, $i=0,1$ and $\tau_{z}=\uparrow,\downarrow$. We have in this 
situation a spin-unpolarized hole gas in the QW. Other LLs from states at the C layer are also occupied but they do 
not affect the results we discuss here. At $B \sim 5$ T, as depicted in Fig. \ref{fig3}(b), $hh_{1,\downarrow}^{1}$ 
starts to be depopulated, spin-polarizing the QW hole gas. It depopulates entirely at $B \sim 7$ T when the hole gas
in the QW becomes partially spin-polarized. In this interval, 5-7 T, $hh_{0,\uparrow}^{1}$ and $hh_{0,\downarrow}^{1}$ have 
a significant difference in their magnetic field dependence, with $hh_{0,\downarrow}^{1}$ energy increasing strongly 
with the magnetic field while $hh_{0,\uparrow}^{1}$ shows a weak dependence with it. At $B \sim 7$ T $hh_{1,\uparrow}^{1}$ 
starts to be depopulated and the QW hole gas starts to decrease its spin-polarization until $B \sim 11$ T when 
$hh_{1,\uparrow}^{1}$ is completely emptied and the QW hole gas is spin-unpolarized. In this interval of magnetic field, both 
$hh_{0,\uparrow}^{1}$ and $hh_{0,\downarrow}^{1}$ states show similar weak magnetic field dependence. At $ B \sim 11$ T 
$hh_{0,\downarrow}^{1}$ starts to be depopulate and the hole gas is again spin-polarized. The magnetic field dependence 
of $hh_{0,\uparrow}^{1}$ and $hh_{0,\downarrow}^{1}$ states again differ significantly, repeating the previous pattern. 
The spin-polarization and spin-unpolarization manifests itself by an oscillation in the LL dependence with the magnetic 
field. These oscillations take place each time a QW LL is emptied and a new one starts to be depopulated.  

It should be observed that during this range of magnetic fields 5-11 T, {\it hh} and
{\it lh} LLs associated to the C layer are also changing their occupation
in relation to their maximum occupation but that does not affect the
$hh_{0,\uparrow}^{1}$ and $hh_{0,\downarrow}^{1}$ magnetic field
dependence. Actually, this picture is confirmed by the charge transfer
between the C layer and the QW. Fig. \ref{fig3}(a) shows that the charge transfer
oscillates following the QW LL filling factor. This charge transfer
has its origin in the thermodynamic equilibrium. However, it is not 
significant due to the weak overlap between $hh^{1}$ and $hh^{2}$ 
wave-functions. Clearly, the most important effect is the spin-polarization 
of the hole gas in the QW, a consequence of the charge transfer between LLs 
within the QW.

We now turn our attention to the consequences of these effects on the optical emission. 
Fig. \ref{fig2}(c) illustrates the fundamental energy transition shift, $E_{T}^{\sigma_{+}(\sigma_{-})}(B)$, 
and the non-linear energy shift [$\Delta E_{T}^{\sigma_{+}(\sigma_{-})}$] as a function of the magnetic 
field. We focus our attention on the non-linear energy shifts where the magnetic effects are more clearly 
displayed. We first observe at low magnetic fields a strong non-linear splitting between the two polarized 
emissions. This is a consequence of the {\it sp-d} interaction. At higher magnetic-fields this splitting 
is superimposed by opposed oscillations originated from the QW holes LLs oscillations. We observe that the 
non-linear behavior for the $\sigma_{-}$ transition increases as the QW hole gas starts to be polarized 
reaching a maximum value when the hole gas is spin-polarized. As this polarization starts to decrease, the 
non-linear behavior for $\sigma_{-}$ decreases. The opposite behavior is observed for the $\sigma_{+}$ 
transition. It shows a negative non-linear behavior which roughly follows the same dependence. As a consequence, 
if we look at the transition energy shift, $E_{T}^{\sigma_{+}}$ ($E_{T}^{\sigma_{-}}$) shows a maximum oscillation 
at odd (even) filling factor. Combining all the effects, the non-linear splitting oscillates between 1.2 meV 
($B=3.8$ T) and a maximum of $2$ meV ($B=15$ T). These values are in qualitative agreement and of the same order 
of the non-linear splitting observed experimentally.

Our results suggest that the oscillations observed experimentally by Gazoto \textit{et al.} 
[Ref. \onlinecite{Gazoto2011}] are a consequence of a combined effect of the spin polarized hole 
gas, the Coulomb exchange interaction and the \textit{p-d} exchange interaction. However, in order 
to observe similar Mn dependence in the effect we consider the Mn ions closer to the QW 
as compared to their samples.
\subsection{Role of interactions and Mn position on the electronic structure}
\noindent 

We shall now investigate how the spin-dependent interactions and Mn position affects 
the electronic properties of the 2DHG. We start our discussion considering the same sample 
parameters as discussed above, but turning off the {\it sp-d} interaction, with Mn ions 
acting as non-magnetic acceptor impurity. This give us a clear picture of the effects induced 
by Coulomb exchange interaction on the hole gas, and consequently, by comparison with the 
previous section, we extract the effects of Mn on the hole gas. In the sequence, we change 
the Mn-doping position to 3 nm and turn-off the Coulomb exchange interaction. This allows us 
to infer about the effects Mn position on the electronic structure. Here we will focus on 
the hole gas LLs and transition energy behavior with the magnetic field. The CB electronic states 
have the same quantitative behavior as in the case described in the previous section.
\begin{figure}[!htb]
\begin{center}
\includegraphics[width=0.5\textwidth]{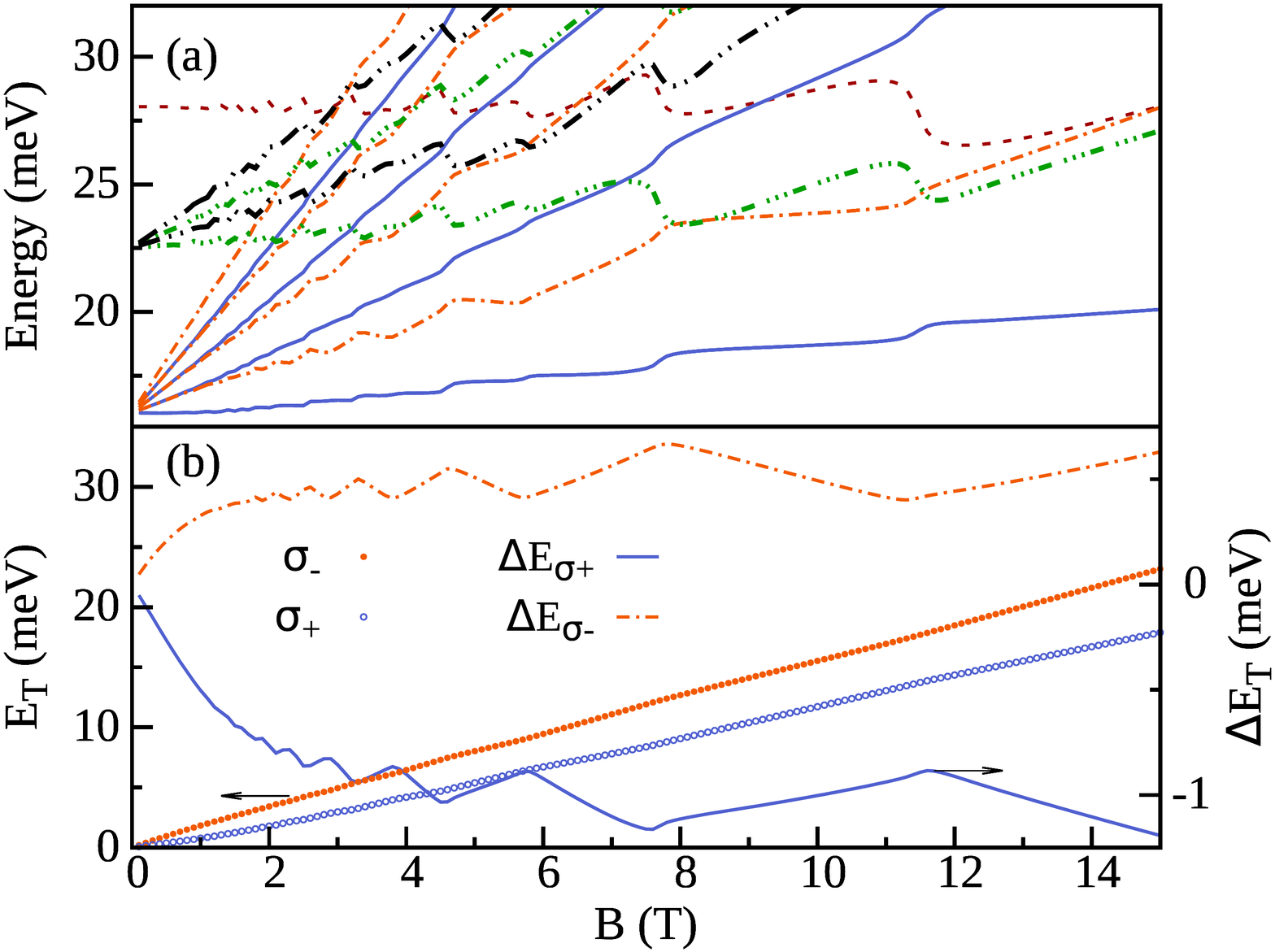}
\caption{(Color online)(a) Heavy-hole Landau Level fan diagram. Solid, dash-dot, dash-three-dots, dash-two-dots, and dashline represents $h_{n,\uparrow}^{1}$, $h_{n,\downarrow,n}^{1}$, $h_{n,\uparrow,n}^{2}$, $h_{n,\downarrow}^{2}$ and $E_{\rm F}$, respectively. (b) Transition energy (right $y$-axis) and non-linear energy shift (left $y$-axis) as a function of the magnetic field. We considered $L_{s}=1$ nm and $V_{pd}^{hh(lh)}=V_{sd}^{e}=0$.\label{fig4}}
\end{center}
\end{figure}

Figure \ref{fig4}(a) illustrates the {\it hh}-LLs fan diagram in absence of the {\it p-d} exchange 
coupling. We first observe that the LLs always increase with the magnetic field independent of the 
{\it hh}-spin, differently of what was observed with the full Hamiltonian for the same heterostructure, 
but in the presence of the hole-Mn spin interaction. The transition energies, shown in Fig. \ref{fig4}(b), 
for both polarizations, increases linearly with the magnetic field, as expected, and show oscillations 
that are directly related with the LLs filling factor. A clear picture of the oscillations is given 
in the non-linear energy shift. For low-magnetic fields ($B<2$ T) the non-linear energy shift presents 
a week dependence with $B$. For higher-magnetic fields it oscillates with LLs filling factor, 
with the maximum (minimum) shift at odd (even) filling factors. This result indicates that the dominant effect 
of the \textit{sp-d} interactions at low magnetic fields with the consequent energy splitting between the different 
polarizations while at higher magnetic fields the spin-polarized hole gas occupation is the dominant effect being responsible 
for the alternate oscillations observed.

\begin{figure}[!htb]
\begin{center} 
\includegraphics[width=0.5\textwidth]{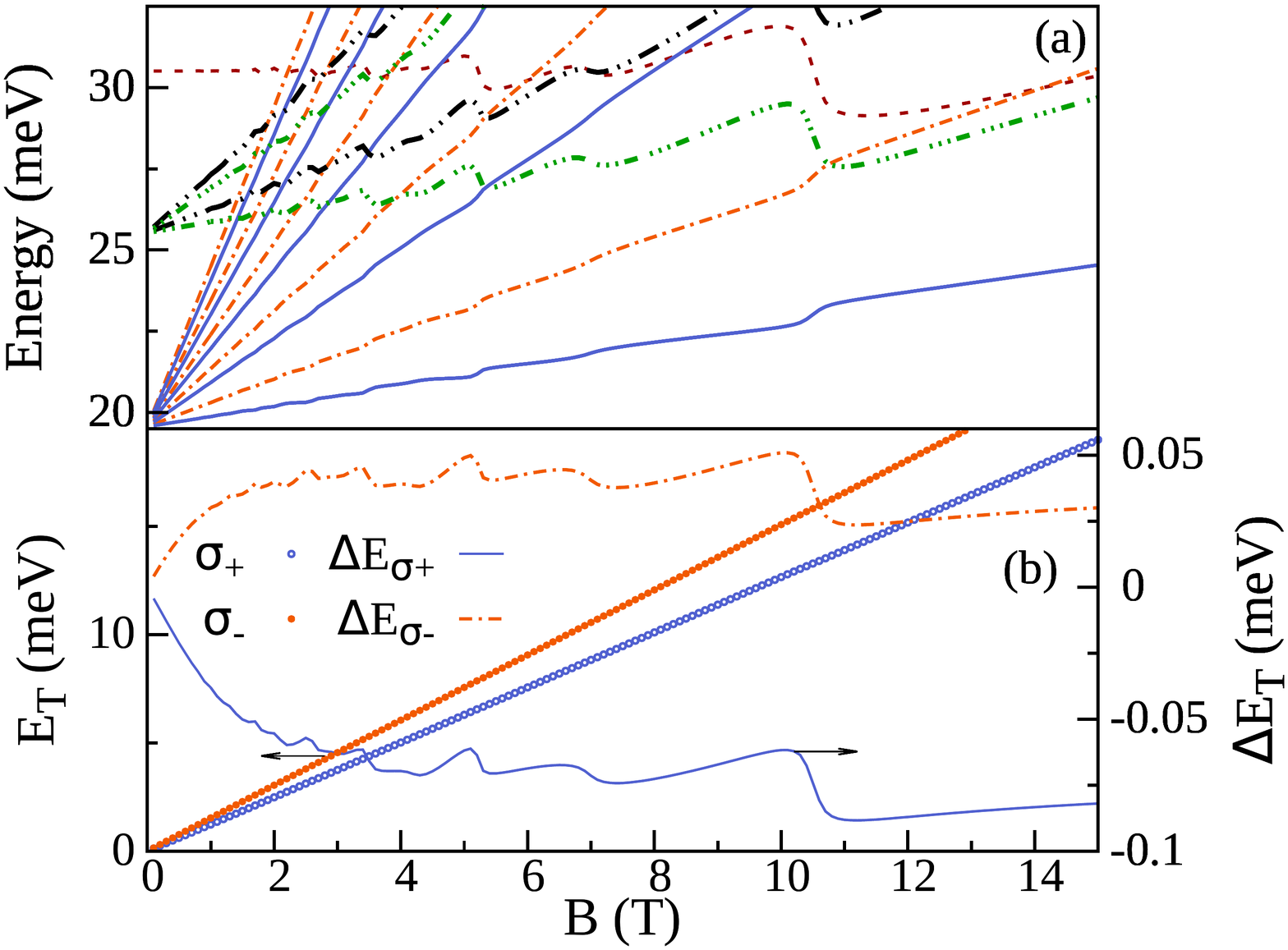} 
\caption{(Color online) (a) LL fan diagram for heavy-holes. The energy states are, in order of increasing energy, labelled as 
$h_{n,\uparrow}^{1}$ (solid), $h_{n,\downarrow,n}^{1}$ (dash-dot), $h_{n,\uparrow,n}^{2}$ (dash-three-dots), 
$h_{n,\downarrow}^{2}$ (dash two-dots) and $E_{\rm F}$(dash-line) , respectively (b) Transition
energy (right $y$-axis) and non-linear energy shift (left $y$-axis)
as function of the magnetic field. We considered $L_{s}=3$ nm and $v_{XC}=0$. \label{fig5}}
\end{center}
\end{figure}

Following, we consider the Mn layer located at 3 nm from the QW/barrier interface. This is the nominal position 
of the Mn layer in the samples investigated in Ref. [\onlinecite{Gazoto2011}]. All the other parameters are the 
same as discussed before. We set now $v_{xc}=0$ in the Hamiltonian, Eq. (\ref{HH_ham}), to focus on the remaining 
\textit{sp-d} interactions. Fig.~\ref{fig5}(a) shows the holes LLs fan diagram.
We observe the non-linear energy splitting between the two polarizations at low magnetic fields. This splitting, however, 
is considerable diminished as a consequence of the weaker overlap of the hole wave-functions with the Mn layer. 
We also observe an oscillatory behavior in the $hh_{n,\tau_{z}}^{1}$ LLs with the magnetic field which is associated 
to the QW LL filling factor. These oscillations, however, have a completely different qualitative and quantitative behavior in 
comparison with the results shown in the previously section. First of all, we do not observe a qualitative dependence 
with the hole-spin in the LLs oscillations. Both fundamental states, $hh_{0,\uparrow}^{1}$ and $hh_{0,\downarrow}^{1}$, 
oscillate following the same pattern with the magnetic field. Second, the value of these oscillations is significantly 
diminished. This can be better visualized in Fig. \ref{fig5}(b) where we plot the energy transition and the non-linear 
energy shift as a function of the magnetic field. The non-linear behavior is the same for both circularly polarized
transitions, as expected from the {\it hh} LLs behavior. The tiny split observed in the total energy shift is of 
the order of $0.1$ meV. These results confirm the combined origin of the oscillations in the Coulomb exchange interaction 
and the \textit{p-d} exchange interaction in the presence of the hole gas. They also demonstrate that 
the Mn ions have to be close to the QW states to observe a sizeable \textit{p-d} coupling. 
\subsection{Gate Voltage}

It is interesting to consider now the possibility to control the magnetic 
effects in the structure by the application of an electric field. This can be 
achieved through a gate voltage modifying the Fermi level in the structure and, 
therefore, the carrier distribution. Essentially, we will simple change the Fermi level at the
surface by the expression $E_{F}\rightarrow E_{g}/2-V_{g}$ and will
consider the value $V_{g}=0.71$ eV which leads to an almost flat
band condition near the surface. We consider a value of $1$ nm for the Mn layer spacing. 
All the other parameters are the same. 
 
\begin{figure}[!htb]
\begin{center}
\includegraphics[width=0.5\textwidth]{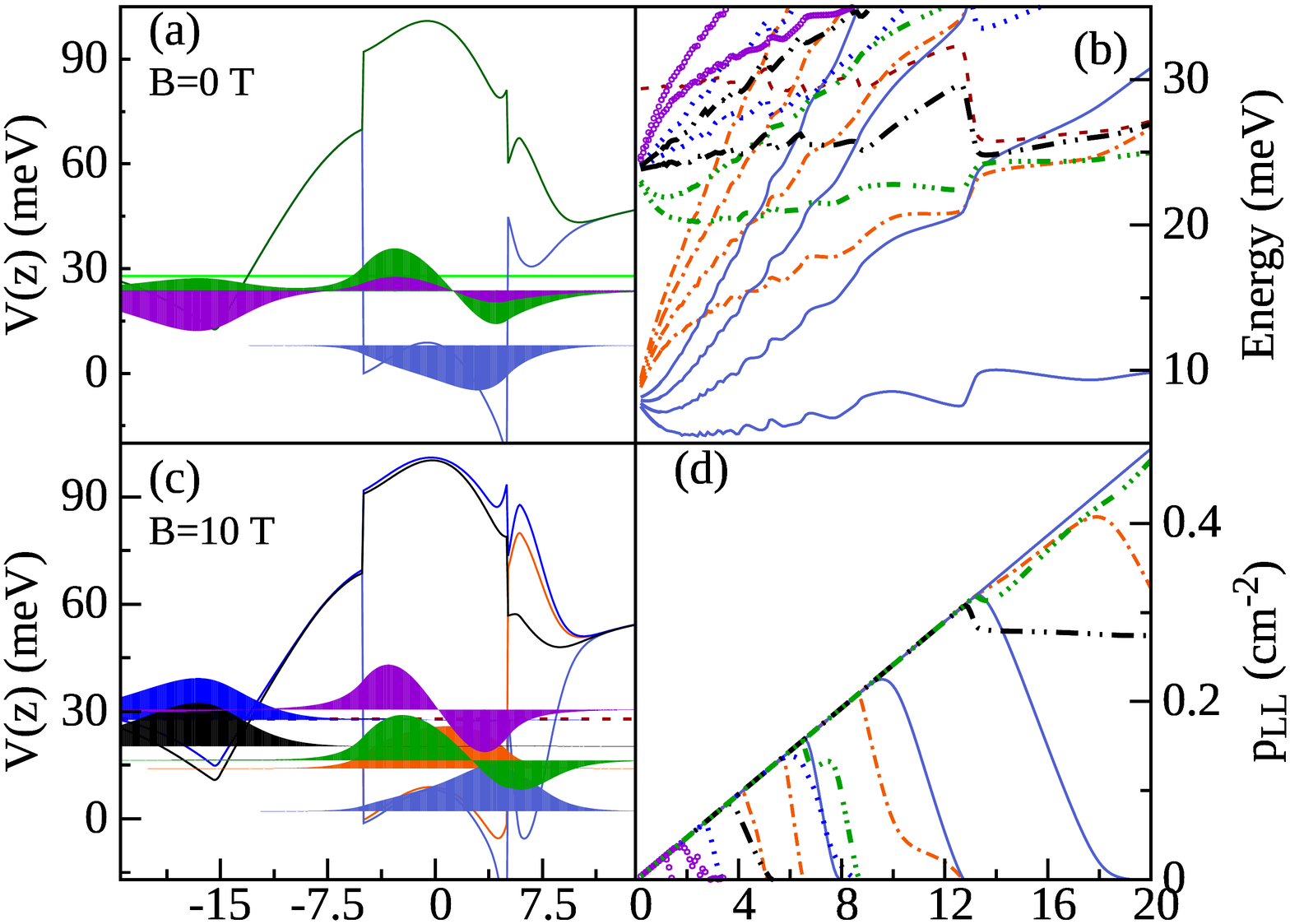} 
\caption{(Color online) (a) and (c) Self-consistent profile potential and
heavy-hole wave-functions for $B=0$ T and $B=10$ T, respectively.
(b) Heavy-hole LLs and (d) two-dimensional density of the LLs. 
The LLs and its density are, in order of increasing energy, labelled as 
$h_{n,\uparrow}^{1}$ (solid), $h_{n,\downarrow,n}^{1}$ (dash-dot), $h_{n,\uparrow,n}^{2}$ (dash-three-dots), 
$h_{n,\downarrow}^{2}$ (dash two-dots) and $E_{\rm F}$(dash-line), respectively. 
For $x_{Mn}=0.4$ MLs, $V_{g}=0.71$ eV and $L_{s}=1$ nm. \label{fig6}}
\end{center}
\end{figure}

Figure \ref{fig6}(a) shows the {\it hh} potential profile and the wave-functions 
of the occupied {\it hh} levels at (a) B=0 T and for (c) B=10 T, (b) the {\it hh} 
LLs fan diagram and (d) the {\it hh}-LLs charge concentration as a function of the 
magnetic field. We first observe that there are three occupied {\it hh} states at $B=0$ T. 
As the magnetic field increases, this picture changes completely. The Mn rich region 
becomes attractive for the spin up {\it hh}s and repulsive for spin down {\it hh}s, 
and the QW spin up levels are pushed towards this region. As a consequence, following 
the previous terminology, $hh_{\uparrow}^{1}$ strongly overlaps with the Mn region.
The same happens for $hh_{\downarrow}^{1}$ although it is less pronounced. 
$hh_{\uparrow}^{2}$ also presents a significant overlap with the Mn region, while 
$hh_{\downarrow}^{2}$ is mainly confined in the QW. The states $hh_{\uparrow}^{3}$ 
and $hh_{\downarrow}^{3}$ become almost fully confined in the C layer. All this 
evolution in the wave-functions produces a complex structure in the LLs fan 
diagram.

We focus now our analysis in the {\it hh} ground-state LLs, that is, 
$hh_{0,\uparrow}^{1}$ and $hh_{0,\downarrow}^{1}$. At low magnetic fields, 
the two states show a strong non-linear behavior, with $hh_{0,\uparrow}^{1}$ 
showing a convex curvature while $hh_{0,\downarrow}^{1}$ shows a concave 
curvature with the magnetic field. This regime is entirely dominated 
by the $p$-$d$ interaction. At higher magnetic fields, for $B>6$ T, 
the oscillations with the LLs filling factors become more pronounced 
and dominate the features. However, in this case, we do not observe the 
difference in the magnetic field dependence from spin up and spin down 
with the odd and even filling factors which was originated from the hole 
gas exchange energy in the presence of the spin-polarization. Here, both 
states oscillate in a similar way. The main reason for this behavior is 
in the way that the LLs associated to the QW are depopulated. This can be 
observed in Fig. \ref{fig6}(d). In particular, for $B\sim13$ T, both
LLs associated to spin up and down become depopulated almost at the
same magnetic field. This prevents the spin-polarization of the QW hole gas.
At the same time, the participation of the states $hh_{\uparrow}^{2}$
in the polarization of the QW hole gas prevents a clear oscillation
in this polarization. On the other way, the states $hh_{0,\tau_{z}}^{1}$
have a strong influence of the Mn ions which dominates the magnetic
field dependence. The oscillations are therefore mainly dominated
by the charge transfer among the QW levels but not in the polarization
of the hole gas.
\begin{figure}[!htb]
\begin{center}
\includegraphics[width=0.5\textwidth]{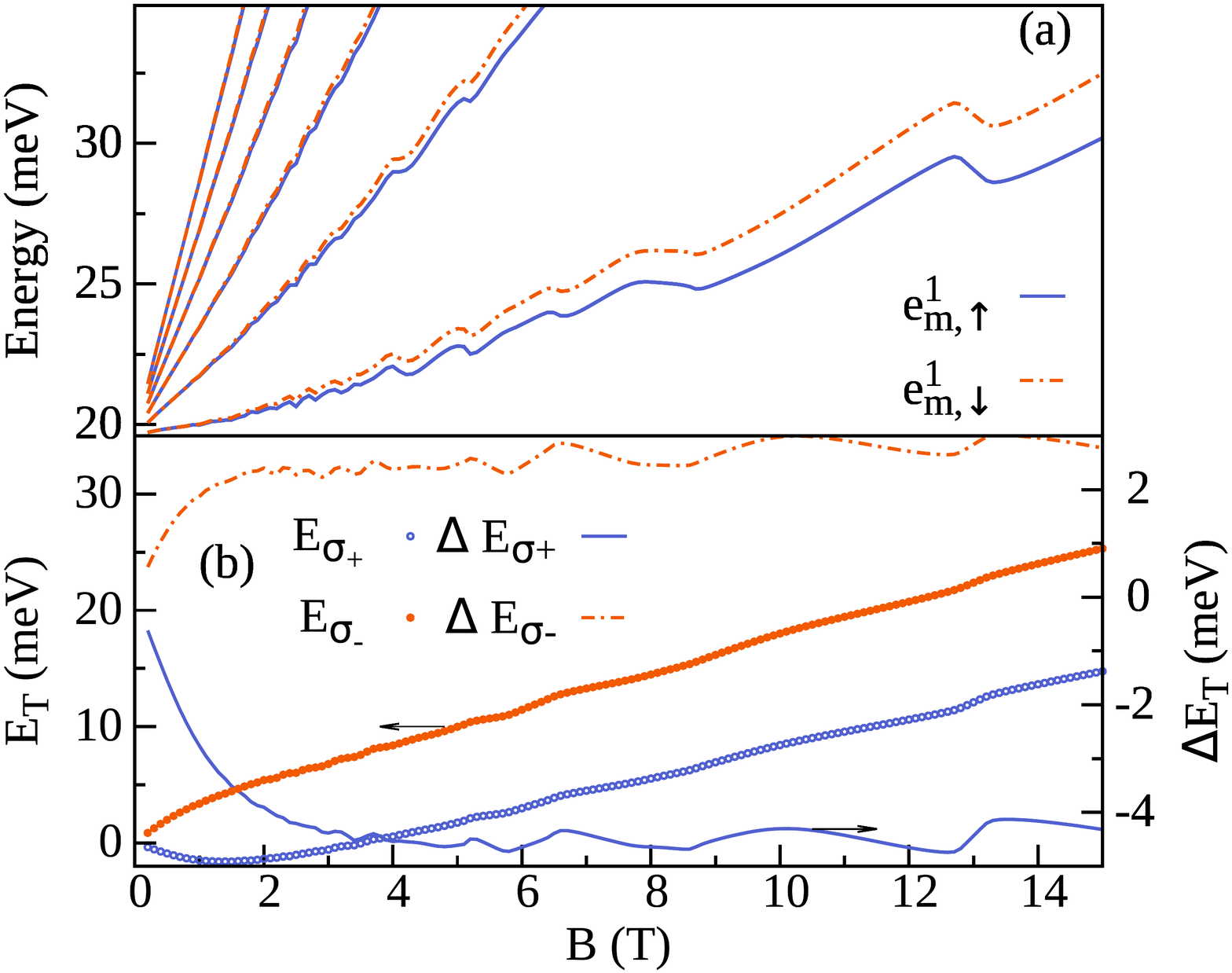} 
\caption{(Color online) (a) CB LL fan diagram . (b) Transition energy (right
$y$-axis) and non-linear energy shift (left $y$-axis) as function
of the magnetic field. $x_{Mn}=0.4$ MLs, $V_{g}=0.71$ eV and $L_{s}=1$
nm. \label{fig7}}
\end{center}
\end{figure}

Figure \ref{fig7} shows (a) the CB LLs fan diagram and (b) the transition
energies and the non-linear energy shift for this case. The CB LLs
show the usual oscillation associated to the charge transfer between
different hole gas reservoirs. The non-linear shift observes a rapid
split at low magnetic fields due to the increasing of the $sp$-$d$
interaction with the magnetic field. For $B>3$ T this splitting
saturates and the oscillatory behavior dominates its features. As it 
was already mentioned for the {\it hh} LLs, we observe symmetric 
oscillations for the two circularly polarized transitions associated to 
the LLs filling factor. These results show that by applying a gate voltage 
we are able to qualitatively change the behaviour of the optical emission.
\section{Concluding Remarks \label{conclusao}}
\noindent 

We presented here the results of calculations of the electronic structure and optical emission energy 
for (In,Ga)As QWs with barriers doped with Mn and C acceptors with a focus on the influence of the Mn 
impurities. Our results clearly show that for a significant magnetic effect on the optical emission the 
Mn layer has to be near the QW interface, allowing a strong overlap between QW wave functions and Mn spin. 
We can separate two regimes for the optical emission effects. At low-magnetic fields, $B<5$ T, 
the {\it p-d} exchange interaction dominates the effects while at higher magnetic fields the spin-polarization 
of the hole gas becomes the dominant effect. We also demonstrated that the non-linear effects can be 
controlled applying an external electric field.  

Our results explain the experimental observations by Gazoto {\it et al.} [Ref. \onlinecite{Gazoto2011}]. The oscillatory 
magneto-shift observed in the magneto-photoluminescence can be interpreted as a combined effect of the exchange hole 
interaction in the presence of the spin-polarized hole gas enhanced by the {\it p-d} interaction. However, to observe a clear 
dependence with the Mn it is necessary to consider a Mn layer closer to the QW than the nominal value.

\section*{ACKNOWLEDGMENT}
We thank F. Iikawa, P. Hawrylak, K. Capelle, P. M. Koenraad and P. A. Bobbert for fruitful discussions. 
The authors acknowledges the support from CAPES-Brazil (Project Number 5860/11-3) and FAPESP-Brazil (Project Number 2010/11393-5).


\begin{thebibliography}{57}
\expandafter\ifx\csname natexlab\endcsname\relax\def\natexlab#1{#1}\fi
\expandafter\ifx\csname bibnamefont\endcsname\relax
  \def\bibnamefont#1{#1}\fi
\expandafter\ifx\csname bibfnamefont\endcsname\relax
  \def\bibfnamefont#1{#1}\fi
\expandafter\ifx\csname citenamefont\endcsname\relax
  \def\citenamefont#1{#1}\fi
\expandafter\ifx\csname url\endcsname\relax
  \def\url#1{\texttt{#1}}\fi
\expandafter\ifx\csname urlprefix\endcsname\relax\def\urlprefix{URL }\fi
\providecommand{\bibinfo}[2]{#2}
\providecommand{\eprint}[2][]{\url{#2}}

\bibitem[{\citenamefont{Munekata et~al.}(1989)\citenamefont{Munekata, Ohno, von
  Molnar, Segm{\"{u}}ller, Chang, and Esaki}}]{Munekata1989}
\bibinfo{author}{\bibfnamefont{H.}~\bibnamefont{Munekata}},
  \bibinfo{author}{\bibfnamefont{H.}~\bibnamefont{Ohno}},
  \bibinfo{author}{\bibfnamefont{S.}~\bibnamefont{von Molnar}},
  \bibinfo{author}{\bibfnamefont{A.}~\bibnamefont{Segm{\"{u}}ller}},
  \bibinfo{author}{\bibfnamefont{L.~L.} \bibnamefont{Chang}}, \bibnamefont{and}
  \bibinfo{author}{\bibfnamefont{L.}~\bibnamefont{Esaki}},
  \bibinfo{journal}{Phys. Rev. Lett.} \textbf{\bibinfo{volume}{63}},
  \bibinfo{pages}{1849} (\bibinfo{year}{1989}).

\bibitem[{\citenamefont{Dietl et~al.}(2000)\citenamefont{Dietl, Ohno,
  Matsukura, Cibert, and Ferrand}}]{Dietl2000}
\bibinfo{author}{\bibfnamefont{T.}~\bibnamefont{Dietl}},
  \bibinfo{author}{\bibfnamefont{H.}~\bibnamefont{Ohno}},
  \bibinfo{author}{\bibfnamefont{F.}~\bibnamefont{Matsukura}},
  \bibinfo{author}{\bibfnamefont{J.}~\bibnamefont{Cibert}}, \bibnamefont{and}
  \bibinfo{author}{\bibfnamefont{D.}~\bibnamefont{Ferrand}},
  \bibinfo{journal}{Science} \textbf{\bibinfo{volume}{287}},
  \bibinfo{pages}{1019} (\bibinfo{year}{2000}).

\bibitem[{\citenamefont{Jungwirth et~al.}(2014)\citenamefont{Jungwirth,
  Wunderlich, Nov{\'{a}}k, Olejn{\'{i}}k, Gallagher, Campion, Edmonds,
  Rushforth, Ferguson, and N{\^{e}}mec}}]{Jungwirth2013a}
\bibinfo{author}{\bibfnamefont{T.}~\bibnamefont{Jungwirth}},
  \bibinfo{author}{\bibfnamefont{J.}~\bibnamefont{Wunderlich}},
  \bibinfo{author}{\bibfnamefont{V.}~\bibnamefont{Nov{\'{a}}k}},
  \bibinfo{author}{\bibfnamefont{K.}~\bibnamefont{Olejn{\'{i}}k}},
  \bibinfo{author}{\bibfnamefont{B.~L.} \bibnamefont{Gallagher}},
  \bibinfo{author}{\bibfnamefont{R.~P.} \bibnamefont{Campion}},
  \bibinfo{author}{\bibfnamefont{K.~W.} \bibnamefont{Edmonds}},
  \bibinfo{author}{\bibfnamefont{A.~W.} \bibnamefont{Rushforth}},
  \bibinfo{author}{\bibfnamefont{A.~J.} \bibnamefont{Ferguson}},
  \bibnamefont{and}
  \bibinfo{author}{\bibfnamefont{P.}~\bibnamefont{N{\^{e}}mec}},
  \bibinfo{journal}{Rev. Mod. Phys.} \textbf{\bibinfo{volume}{86}},
  \bibinfo{pages}{855} (\bibinfo{year}{2014}).

\bibitem[{\citenamefont{Olejn{\'{i}}k et~al.}(2008)\citenamefont{Olejn{\'{i}}k,
  Owen, Nov{\'{a}}k, Ma{\^{s}}ek, Irvine, Wunderlich, and
  Jungwirth}}]{Olejnik2008}
\bibinfo{author}{\bibfnamefont{K.}~\bibnamefont{Olejn{\'{i}}k}},
  \bibinfo{author}{\bibfnamefont{M.~H.~S.} \bibnamefont{Owen}},
  \bibinfo{author}{\bibfnamefont{V.}~\bibnamefont{Nov{\'{a}}k}},
  \bibinfo{author}{\bibfnamefont{J.}~\bibnamefont{Ma{\^{s}}ek}},
  \bibinfo{author}{\bibfnamefont{A.~C.} \bibnamefont{Irvine}},
  \bibinfo{author}{\bibfnamefont{J.}~\bibnamefont{Wunderlich}},
  \bibnamefont{and}
  \bibinfo{author}{\bibfnamefont{T.}~\bibnamefont{Jungwirth}},
  \bibinfo{journal}{Phys. Rev. B} \textbf{\bibinfo{volume}{78}},
  \bibinfo{pages}{054403} (\bibinfo{year}{2008}).

\bibitem[{\citenamefont{Chen et~al.}(2009)\citenamefont{Chen, Yan, Xu, Lu,
  Wang, Deng, Qian, Ji, and Zhaoa}}]{Chen2009}
\bibinfo{author}{\bibfnamefont{L.}~\bibnamefont{Chen}},
  \bibinfo{author}{\bibfnamefont{S.}~\bibnamefont{Yan}},
  \bibinfo{author}{\bibfnamefont{P.~F.} \bibnamefont{Xu}},
  \bibinfo{author}{\bibfnamefont{J.}~\bibnamefont{Lu}},
  \bibinfo{author}{\bibfnamefont{W.~Z.} \bibnamefont{Wang}},
  \bibinfo{author}{\bibfnamefont{J.~J.} \bibnamefont{Deng}},
  \bibinfo{author}{\bibfnamefont{X.}~\bibnamefont{Qian}},
  \bibinfo{author}{\bibfnamefont{Y.}~\bibnamefont{Ji}}, \bibnamefont{and}
  \bibinfo{author}{\bibfnamefont{J.~H.} \bibnamefont{Zhaoa}},
  \bibinfo{journal}{Appl. Phys. Lett.} \textbf{\bibinfo{volume}{95}},
  \bibinfo{pages}{182505} (\bibinfo{year}{2009}).

\bibitem[{\citenamefont{Dietl and Ohno}(2014)}]{dietl-ohno-rmp2014}
\bibinfo{author}{\bibfnamefont{T.}~\bibnamefont{Dietl}} \bibnamefont{and}
  \bibinfo{author}{\bibfnamefont{H.}~\bibnamefont{Ohno}},
  \bibinfo{journal}{Rev. Mod. Phys.} \textbf{\bibinfo{volume}{86}},
  \bibinfo{pages}{187} (\bibinfo{year}{2014}).

\bibitem[{\citenamefont{Jungwirth et~al.}(2006)\citenamefont{Jungwirth, Sinova,
  Ma{\v{s}}ek, Ku{\v{c}}era, and MacDonald}}]{Jungwirth2006}
\bibinfo{author}{\bibfnamefont{T.}~\bibnamefont{Jungwirth}},
  \bibinfo{author}{\bibfnamefont{J.}~\bibnamefont{Sinova}},
  \bibinfo{author}{\bibfnamefont{J.}~\bibnamefont{Ma{\v{s}}ek}},
  \bibinfo{author}{\bibfnamefont{J.}~\bibnamefont{Ku{\v{c}}era}},
  \bibnamefont{and} \bibinfo{author}{\bibfnamefont{A.~H.}
  \bibnamefont{MacDonald}}, \bibinfo{journal}{Rev. Mod. Phys.}
  \textbf{\bibinfo{volume}{78}}, \bibinfo{pages}{809} (\bibinfo{year}{2006}).

\bibitem[{\citenamefont{Chiba et~al.}(2008)\citenamefont{Chiba, Sawicki,
  Nishitani, Nakatani, Matsukura, and Ohno}}]{Chiba2008}
\bibinfo{author}{\bibfnamefont{D.}~\bibnamefont{Chiba}},
  \bibinfo{author}{\bibfnamefont{M.}~\bibnamefont{Sawicki}},
  \bibinfo{author}{\bibfnamefont{Y.}~\bibnamefont{Nishitani}},
  \bibinfo{author}{\bibfnamefont{Y.}~\bibnamefont{Nakatani}},
  \bibinfo{author}{\bibfnamefont{F.}~\bibnamefont{Matsukura}},
  \bibnamefont{and} \bibinfo{author}{\bibfnamefont{H.}~\bibnamefont{Ohno}},
  \bibinfo{journal}{Nature} \textbf{\bibinfo{volume}{455}},
  \bibinfo{pages}{515} (\bibinfo{year}{2008}).

\bibitem[{\citenamefont{Nemec et~al.}(2012)\citenamefont{Nemec,
  Rozkotov{\'{a}}, Tesarov{\'{a}}, Troj{\'{a}}nek, Ranieri, Olejn{\'{i}}k,
  Zemen, Nov{\'{a}}k, Cukr, Mal{\'{y}} et~al.}}]{Nemec2012}
\bibinfo{author}{\bibfnamefont{P.}~\bibnamefont{Nemec}},
  \bibinfo{author}{\bibfnamefont{E.}~\bibnamefont{Rozkotov{\'{a}}}},
  \bibinfo{author}{\bibfnamefont{N.}~\bibnamefont{Tesarov{\'{a}}}},
  \bibinfo{author}{\bibfnamefont{F.}~\bibnamefont{Troj{\'{a}}nek}},
  \bibinfo{author}{\bibfnamefont{E.~D.} \bibnamefont{Ranieri}},
  \bibinfo{author}{\bibfnamefont{K.}~\bibnamefont{Olejn{\'{i}}k}},
  \bibinfo{author}{\bibfnamefont{J.}~\bibnamefont{Zemen}},
  \bibinfo{author}{\bibfnamefont{V.}~\bibnamefont{Nov{\'{a}}k}},
  \bibinfo{author}{\bibfnamefont{M.}~\bibnamefont{Cukr}},
  \bibinfo{author}{\bibfnamefont{P.}~\bibnamefont{Mal{\'{y}}}},
  \bibnamefont{et~al.}, \bibinfo{journal}{Nat. Phys.}
  \textbf{\bibinfo{volume}{8}}, \bibinfo{pages}{411} (\bibinfo{year}{2012}).

\bibitem[{\citenamefont{Tesa{\^{r}}ov{\'{a}}
  et~al.}(2013)\citenamefont{Tesa{\^{r}}ov{\'{a}}, N{\^{e}}mec,
  Rozkotov{\'{a}}, Zemen, Janda, D.~Butkovi{\^{c}}ova´1, Olejn{\'{i}}k,
  Nov{\'{a}}k, Mal{\'{y}}, and Jungwirth}}]{Tesarova2013}
\bibinfo{author}{\bibfnamefont{N.}~\bibnamefont{Tesa{\^{r}}ov{\'{a}}}},
  \bibinfo{author}{\bibfnamefont{P.}~\bibnamefont{N{\^{e}}mec}},
  \bibinfo{author}{\bibfnamefont{E.}~\bibnamefont{Rozkotov{\'{a}}}},
  \bibinfo{author}{\bibfnamefont{J.}~\bibnamefont{Zemen}},
  \bibinfo{author}{\bibfnamefont{T.}~\bibnamefont{Janda}},
  \bibinfo{author}{\bibfnamefont{F.~T.} \bibnamefont{D.~Butkovi{\^{c}}ova´1}},
  \bibinfo{author}{\bibfnamefont{K.}~\bibnamefont{Olejn{\'{i}}k}},
  \bibinfo{author}{\bibfnamefont{V.}~\bibnamefont{Nov{\'{a}}k}},
  \bibinfo{author}{\bibfnamefont{P.}~\bibnamefont{Mal{\'{y}}}},
  \bibnamefont{and}
  \bibinfo{author}{\bibfnamefont{T.}~\bibnamefont{Jungwirth}},
  \bibinfo{journal}{Nat. Photon.} \textbf{\bibinfo{volume}{7}},
  \bibinfo{pages}{492} (\bibinfo{year}{2013}).

\bibitem[{\citenamefont{Wurstbauer et~al.}(2010)\citenamefont{Wurstbauer,
  {\'{S}}liwa, Weiss, Dietl, and Wegscheider}}]{Wurstbauer2010}
\bibinfo{author}{\bibfnamefont{U.}~\bibnamefont{Wurstbauer}},
  \bibinfo{author}{\bibfnamefont{C.}~\bibnamefont{{\'{S}}liwa}},
  \bibinfo{author}{\bibfnamefont{D.}~\bibnamefont{Weiss}},
  \bibinfo{author}{\bibfnamefont{T.}~\bibnamefont{Dietl}}, \bibnamefont{and}
  \bibinfo{author}{\bibfnamefont{W.}~\bibnamefont{Wegscheider}},
  \bibinfo{journal}{Nat. Phys.} \textbf{\bibinfo{volume}{6}},
  \bibinfo{pages}{955} (\bibinfo{year}{2010}).

\bibitem[{\citenamefont{Knott et~al.}(2011)\citenamefont{Knott, Hirschmann,
  Wurstbauer, Hansen, and Wegscheider}}]{Knott2011}
\bibinfo{author}{\bibfnamefont{S.}~\bibnamefont{Knott}},
  \bibinfo{author}{\bibfnamefont{T.~C.} \bibnamefont{Hirschmann}},
  \bibinfo{author}{\bibfnamefont{U.}~\bibnamefont{Wurstbauer}},
  \bibinfo{author}{\bibfnamefont{W.}~\bibnamefont{Hansen}}, \bibnamefont{and}
  \bibinfo{author}{\bibfnamefont{W.}~\bibnamefont{Wegscheider}},
  \bibinfo{journal}{Phys. Rev. B} \textbf{\bibinfo{volume}{84}},
  \bibinfo{pages}{205302} (\bibinfo{year}{2011}).

\bibitem[{\citenamefont{Zaitsev and Zvonkov}(2010)}]{Zaitsev2010}
\bibinfo{author}{\bibfnamefont{S.~V.} \bibnamefont{Zaitsev}} \bibnamefont{and}
  \bibinfo{author}{\bibfnamefont{B.~N.} \bibnamefont{Zvonkov}},
  \bibinfo{journal}{Phys. Status Solidi B} \textbf{\bibinfo{volume}{248}},
  \bibinfo{pages}{1526} (\bibinfo{year}{2010}).

\bibitem[{\citenamefont{Gazoto et~al.}(2011)\citenamefont{Gazoto, Brasil,
  Iikawa, Brum, Ribeiro, Danilov, Vikhrova, and Zvonkov}}]{Gazoto2011}
\bibinfo{author}{\bibfnamefont{A.~L.} \bibnamefont{Gazoto}},
  \bibinfo{author}{\bibfnamefont{M.~J. S.~P.} \bibnamefont{Brasil}},
  \bibinfo{author}{\bibfnamefont{F.}~\bibnamefont{Iikawa}},
  \bibinfo{author}{\bibfnamefont{J.~A.} \bibnamefont{Brum}},
  \bibinfo{author}{\bibfnamefont{E.}~\bibnamefont{Ribeiro}},
  \bibinfo{author}{\bibfnamefont{Y.~A.} \bibnamefont{Danilov}},
  \bibinfo{author}{\bibfnamefont{O.~V.} \bibnamefont{Vikhrova}},
  \bibnamefont{and} \bibinfo{author}{\bibfnamefont{B.~N.}
  \bibnamefont{Zvonkov}}, \bibinfo{journal}{Appl. Phys. Lett.}
  \textbf{\bibinfo{volume}{98}}, \bibinfo{pages}{251901}
  (\bibinfo{year}{2011}).

\bibitem[{\citenamefont{Korenev et~al.}(2012)\citenamefont{Korenev, Akimov,
  Zaitsev, Sapega, Langer, and. Yu. A.~Danilov, and Bayer}}]{Korenev2012}
\bibinfo{author}{\bibfnamefont{V.~L.} \bibnamefont{Korenev}},
  \bibinfo{author}{\bibfnamefont{I.~A.} \bibnamefont{Akimov}},
  \bibinfo{author}{\bibfnamefont{S.~V.} \bibnamefont{Zaitsev}},
  \bibinfo{author}{\bibfnamefont{V.~F.} \bibnamefont{Sapega}},
  \bibinfo{author}{\bibfnamefont{L.}~\bibnamefont{Langer}},
  \bibinfo{author}{\bibfnamefont{D.~R.~Y.} \bibnamefont{and. Yu. A.~Danilov}},
  \bibnamefont{and} \bibinfo{author}{\bibfnamefont{M.}~\bibnamefont{Bayer}},
  \bibinfo{journal}{Nat. Commun.} \textbf{\bibinfo{volume}{3}},
  \bibinfo{pages}{959} (\bibinfo{year}{2012}).

\bibitem[{\citenamefont{Balanta et~al.}(2013)\citenamefont{Balanta, Brasil,
  Iikawa, Mendes, Brum, Maialle, Danilov, Vikhrova, and Zvonkov}}]{Balanta2013}
\bibinfo{author}{\bibfnamefont{M.~A.~G.} \bibnamefont{Balanta}},
  \bibinfo{author}{\bibfnamefont{M.~J. S.~P.} \bibnamefont{Brasil}},
  \bibinfo{author}{\bibfnamefont{F.}~\bibnamefont{Iikawa}},
  \bibinfo{author}{\bibfnamefont{U.~C.} \bibnamefont{Mendes}},
  \bibinfo{author}{\bibfnamefont{J.~A.} \bibnamefont{Brum}},
  \bibinfo{author}{\bibfnamefont{M.~Z.} \bibnamefont{Maialle}},
  \bibinfo{author}{\bibfnamefont{Y.~A.} \bibnamefont{Danilov}},
  \bibinfo{author}{\bibfnamefont{O.~V.} \bibnamefont{Vikhrova}},
  \bibnamefont{and} \bibinfo{author}{\bibfnamefont{B.~N.}
  \bibnamefont{Zvonkov}}, \bibinfo{journal}{J. Phys. D: Appl. Phys.}
  \textbf{\bibinfo{volume}{46}}, \bibinfo{pages}{215103}
  (\bibinfo{year}{2013}).

\bibitem[{\citenamefont{Balanta et~al.}(2014)\citenamefont{Balanta, Brasil,
  Iikawa, Brum, Mendes, Danilov, Dorokhin, Vikhrova, and
  Zvonkov}}]{balanta-brasil-jap2014}
\bibinfo{author}{\bibfnamefont{M.~A.~G.} \bibnamefont{Balanta}},
  \bibinfo{author}{\bibfnamefont{M.~J. S.~P.} \bibnamefont{Brasil}},
  \bibinfo{author}{\bibfnamefont{F.}~\bibnamefont{Iikawa}},
  \bibinfo{author}{\bibfnamefont{J.~A.} \bibnamefont{Brum}},
  \bibinfo{author}{\bibfnamefont{U.~C.} \bibnamefont{Mendes}},
  \bibinfo{author}{\bibfnamefont{Y.~A.} \bibnamefont{Danilov}},
  \bibinfo{author}{\bibfnamefont{M.~V.} \bibnamefont{Dorokhin}},
  \bibinfo{author}{\bibfnamefont{O.~V.} \bibnamefont{Vikhrova}},
  \bibnamefont{and} \bibinfo{author}{\bibfnamefont{B.~N.}
  \bibnamefont{Zvonkov}}, \bibinfo{journal}{J. Appl. Phys.}
  \textbf{\bibinfo{volume}{116}}, \bibinfo{pages}{203501}
  (\bibinfo{year}{2014}).

\bibitem[{\citenamefont{Kerridge et~al.}(1999)\citenamefont{Kerridge, Greally,
  Hayne, Usher, Plaut, Brum, Holland, and Stanley}}]{kerridge_brum_ssc1999}
\bibinfo{author}{\bibfnamefont{G.~C.} \bibnamefont{Kerridge}},
  \bibinfo{author}{\bibfnamefont{M.~G.} \bibnamefont{Greally}},
  \bibinfo{author}{\bibfnamefont{M.}~\bibnamefont{Hayne}},
  \bibinfo{author}{\bibfnamefont{A.}~\bibnamefont{Usher}},
  \bibinfo{author}{\bibfnamefont{A.~S.} \bibnamefont{Plaut}},
  \bibinfo{author}{\bibfnamefont{J.~A.} \bibnamefont{Brum}},
  \bibinfo{author}{\bibfnamefont{M.~C.} \bibnamefont{Holland}},
  \bibnamefont{and} \bibinfo{author}{\bibfnamefont{C.~R.}
  \bibnamefont{Stanley}}, \bibinfo{journal}{Solid State Commun.}
  \textbf{\bibinfo{volume}{109}}, \bibinfo{pages}{267} (\bibinfo{year}{1999}).

\bibitem[{\citenamefont{Kunc et~al.}(2010)\citenamefont{Kunc, Kowalik, Teran,
  Plochocka, Piot, Maude, Potemski, Kolkovsky, Karczewski, and
  Wojtowicz}}]{Kunc2010}
\bibinfo{author}{\bibfnamefont{J.}~\bibnamefont{Kunc}},
  \bibinfo{author}{\bibfnamefont{K.}~\bibnamefont{Kowalik}},
  \bibinfo{author}{\bibfnamefont{F.~J.} \bibnamefont{Teran}},
  \bibinfo{author}{\bibfnamefont{P.}~\bibnamefont{Plochocka}},
  \bibinfo{author}{\bibfnamefont{B.~A.} \bibnamefont{Piot}},
  \bibinfo{author}{\bibfnamefont{D.~K.} \bibnamefont{Maude}},
  \bibinfo{author}{\bibfnamefont{M.}~\bibnamefont{Potemski}},
  \bibinfo{author}{\bibfnamefont{V.}~\bibnamefont{Kolkovsky}},
  \bibinfo{author}{\bibfnamefont{G.}~\bibnamefont{Karczewski}},
  \bibnamefont{and}
  \bibinfo{author}{\bibfnamefont{T.}~\bibnamefont{Wojtowicz}},
  \bibinfo{journal}{Phys. Rev. B} \textbf{\bibinfo{volume}{82}},
  \bibinfo{pages}{115438} (\bibinfo{year}{2010}).

\bibitem[{\citenamefont{Kehoe et~al.}(2003)\citenamefont{Kehoe, Townsley,
  Usher, Henini, and Hill}}]{Kehoe2003}
\bibinfo{author}{\bibfnamefont{T.~B.} \bibnamefont{Kehoe}},
  \bibinfo{author}{\bibfnamefont{C.~M.} \bibnamefont{Townsley}},
  \bibinfo{author}{\bibfnamefont{A.}~\bibnamefont{Usher}},
  \bibinfo{author}{\bibfnamefont{M.}~\bibnamefont{Henini}}, \bibnamefont{and}
  \bibinfo{author}{\bibfnamefont{G.}~\bibnamefont{Hill}},
  \bibinfo{journal}{Phys. Rev. B} \textbf{\bibinfo{volume}{68}},
  \bibinfo{pages}{045325} (\bibinfo{year}{2003}).

\bibitem[{\citenamefont{Uenoyama and Sham}(1989)}]{Uenoyama1989}
\bibinfo{author}{\bibfnamefont{T.}~\bibnamefont{Uenoyama}} \bibnamefont{and}
  \bibinfo{author}{\bibfnamefont{L.~J.} \bibnamefont{Sham}},
  \bibinfo{journal}{Phys. Rev. B} \textbf{\bibinfo{volume}{39}},
  \bibinfo{pages}{11044} (\bibinfo{year}{1989}).

\bibitem[{\citenamefont{Katayama and Ando}(1989)}]{katayama_ando_ssc1989}
\bibinfo{author}{\bibfnamefont{S.}~\bibnamefont{Katayama}} \bibnamefont{and}
  \bibinfo{author}{\bibfnamefont{T.}~\bibnamefont{Ando}},
  \bibinfo{journal}{Solid State Commun.} \textbf{\bibinfo{volume}{70}},
  \bibinfo{pages}{91} (\bibinfo{year}{1989}).

\bibitem[{\citenamefont{Hawrylak and Potemski}(1997)}]{Hawrylak1997}
\bibinfo{author}{\bibfnamefont{P.}~\bibnamefont{Hawrylak}} \bibnamefont{and}
  \bibinfo{author}{\bibfnamefont{M.}~\bibnamefont{Potemski}},
  \bibinfo{journal}{Phys. Rev. B} \textbf{\bibinfo{volume}{56}},
  \bibinfo{pages}{12386} (\bibinfo{year}{1997}).

\bibitem[{\citenamefont{Takeyama et~al.}(1999)\citenamefont{Takeyama,
  Karczewski, Wojtowicz, Kossut, Kunimatsu, Uchida, and Miura}}]{Takeyama1999}
\bibinfo{author}{\bibfnamefont{S.}~\bibnamefont{Takeyama}},
  \bibinfo{author}{\bibfnamefont{G.}~\bibnamefont{Karczewski}},
  \bibinfo{author}{\bibfnamefont{T.}~\bibnamefont{Wojtowicz}},
  \bibinfo{author}{\bibfnamefont{J.}~\bibnamefont{Kossut}},
  \bibinfo{author}{\bibfnamefont{H.}~\bibnamefont{Kunimatsu}},
  \bibinfo{author}{\bibfnamefont{K.}~\bibnamefont{Uchida}}, \bibnamefont{and}
  \bibinfo{author}{\bibfnamefont{N.}~\bibnamefont{Miura}},
  \bibinfo{journal}{Phys. Rev. B} \textbf{\bibinfo{volume}{59}},
  \bibinfo{pages}{7327} (\bibinfo{year}{1999}).

\bibitem[{\citenamefont{Asano and Ando}(2002)}]{Asano2002}
\bibinfo{author}{\bibfnamefont{K.}~\bibnamefont{Asano}} \bibnamefont{and}
  \bibinfo{author}{\bibfnamefont{T.}~\bibnamefont{Ando}},
  \bibinfo{journal}{Phys. Rev. B} \textbf{\bibinfo{volume}{65}},
  \bibinfo{pages}{115330} (\bibinfo{year}{2002}).

\bibitem[{\citenamefont{Ponomarev et~al.}(1996)\citenamefont{Ponomarev, Usher,
  Rodgers, Gallagher, Henini, and Hill}}]{Ponomarev1996}
\bibinfo{author}{\bibfnamefont{Y.~V.} \bibnamefont{Ponomarev}},
  \bibinfo{author}{\bibfnamefont{A.}~\bibnamefont{Usher}},
  \bibinfo{author}{\bibfnamefont{P.~J.} \bibnamefont{Rodgers}},
  \bibinfo{author}{\bibfnamefont{B.~L.} \bibnamefont{Gallagher}},
  \bibinfo{author}{\bibfnamefont{M.}~\bibnamefont{Henini}}, \bibnamefont{and}
  \bibinfo{author}{\bibfnamefont{G.}~\bibnamefont{Hill}},
  \bibinfo{journal}{Phys. Rev. B} \textbf{\bibinfo{volume}{54}},
  \bibinfo{pages}{13891} (\bibinfo{year}{1996}).

\bibitem[{\citenamefont{Aronzon et~al.}(2010)\citenamefont{Aronzon, Pankov,
  Rylkov, Meilikhov, Lagutin, Pashaev, Chuev, Kvardakov, Likhachev, Vihrova
  et~al.}}]{Aronzon2010}
\bibinfo{author}{\bibfnamefont{B.~A.} \bibnamefont{Aronzon}},
  \bibinfo{author}{\bibfnamefont{M.~A.} \bibnamefont{Pankov}},
  \bibinfo{author}{\bibfnamefont{V.~V.} \bibnamefont{Rylkov}},
  \bibinfo{author}{\bibfnamefont{E.~Z.} \bibnamefont{Meilikhov}},
  \bibinfo{author}{\bibfnamefont{A.~S.} \bibnamefont{Lagutin}},
  \bibinfo{author}{\bibfnamefont{E.~M.} \bibnamefont{Pashaev}},
  \bibinfo{author}{\bibfnamefont{M.~A.} \bibnamefont{Chuev}},
  \bibinfo{author}{\bibfnamefont{V.~V.} \bibnamefont{Kvardakov}},
  \bibinfo{author}{\bibfnamefont{I.~A.} \bibnamefont{Likhachev}},
  \bibinfo{author}{\bibfnamefont{O.~V.} \bibnamefont{Vihrova}},
  \bibnamefont{et~al.}, \bibinfo{journal}{J. Appl. Phys.}
  \textbf{\bibinfo{volume}{107}}, \bibinfo{pages}{023905}
  (\bibinfo{year}{2010}).

\bibitem[{\citenamefont{Hohenberg and Kohn}(1964)}]{Hohenberg1964}
\bibinfo{author}{\bibfnamefont{P.}~\bibnamefont{Hohenberg}} \bibnamefont{and}
  \bibinfo{author}{\bibfnamefont{W.}~\bibnamefont{Kohn}},
  \bibinfo{journal}{Phys. Rev.} \textbf{\bibinfo{volume}{136}},
  \bibinfo{pages}{864} (\bibinfo{year}{1964}).

\bibitem[{\citenamefont{Kohn and Sham}(1965)}]{Kohn1965}
\bibinfo{author}{\bibfnamefont{W.}~\bibnamefont{Kohn}} \bibnamefont{and}
  \bibinfo{author}{\bibfnamefont{L.~J.} \bibnamefont{Sham}},
  \bibinfo{journal}{Phys. Rev.} \textbf{\bibinfo{volume}{140}},
  \bibinfo{pages}{1133} (\bibinfo{year}{1965}).

\bibitem[{\citenamefont{Gunnarsson and Lundqvist}(1976)}]{Gunnarsson1976}
\bibinfo{author}{\bibfnamefont{O.}~\bibnamefont{Gunnarsson}} \bibnamefont{and}
  \bibinfo{author}{\bibfnamefont{B.~I.} \bibnamefont{Lundqvist}},
  \bibinfo{journal}{Phys. Rev. B} \textbf{\bibinfo{volume}{13}},
  \bibinfo{pages}{4274} (\bibinfo{year}{1976}).

\bibitem[{\citenamefont{Gupta and Rajagopal}(1982)}]{Gupta1982}
\bibinfo{author}{\bibfnamefont{U.}~\bibnamefont{Gupta}} \bibnamefont{and}
  \bibinfo{author}{\bibfnamefont{A.~K.} \bibnamefont{Rajagopal}},
  \bibinfo{journal}{Physics Reports} \textbf{\bibinfo{volume}{87}},
  \bibinfo{pages}{259} (\bibinfo{year}{1982}).

\bibitem[{\citenamefont{Bastard}(1992)}]{bastard_book}
\bibinfo{author}{\bibfnamefont{G.}~\bibnamefont{Bastard}},
  \emph{\bibinfo{title}{Wave Mechanics Applied to Semiconductor
  Heterostructures}}, Monographies de physique (\bibinfo{publisher}{Les
  Editions de Physique}, \bibinfo{address}{Paris}, \bibinfo{year}{1992}).

\bibitem[{\citenamefont{Abolfath et~al.}(2001)\citenamefont{Abolfath,
  Jungwirth, Brum, and MacDonald}}]{Abolfath2001}
\bibinfo{author}{\bibfnamefont{M.}~\bibnamefont{Abolfath}},
  \bibinfo{author}{\bibfnamefont{T.}~\bibnamefont{Jungwirth}},
  \bibinfo{author}{\bibfnamefont{J.}~\bibnamefont{Brum}}, \bibnamefont{and}
  \bibinfo{author}{\bibfnamefont{A.~H.} \bibnamefont{MacDonald}},
  \bibinfo{journal}{Phys. Rev. B} \textbf{\bibinfo{volume}{63}},
  \bibinfo{pages}{054418} (\bibinfo{year}{2001}).

\bibitem[{\citenamefont{Luttinger and Kohn}(1955)}]{Luttinger1955}
\bibinfo{author}{\bibfnamefont{J.~M.} \bibnamefont{Luttinger}}
  \bibnamefont{and} \bibinfo{author}{\bibfnamefont{W.}~\bibnamefont{Kohn}},
  \bibinfo{journal}{Phys. Rev.} \textbf{\bibinfo{volume}{97}},
  \bibinfo{pages}{869} (\bibinfo{year}{1955}).

\bibitem[{\citenamefont{von Barth and Hedin}(1972)}]{Barth1972}
\bibinfo{author}{\bibfnamefont{U.}~\bibnamefont{von Barth}} \bibnamefont{and}
  \bibinfo{author}{\bibfnamefont{L.}~\bibnamefont{Hedin}}, \bibinfo{journal}{J.
  Phys. C} \textbf{\bibinfo{volume}{5}}, \bibinfo{pages}{1629}
  (\bibinfo{year}{1972}).

\bibitem[{\citenamefont{Bastard and Brum}(1986)}]{Bastard1986}
\bibinfo{author}{\bibfnamefont{G.}~\bibnamefont{Bastard}} \bibnamefont{and}
  \bibinfo{author}{\bibfnamefont{J.~A.} \bibnamefont{Brum}},
  \bibinfo{journal}{IEEE Journal of Quantum Electronics}
  \textbf{\bibinfo{volume}{22}}, \bibinfo{pages}{1625} (\bibinfo{year}{1986}).

\bibitem[{\citenamefont{Mace et~al.}(1988)\citenamefont{Mace, Rogers,
  Monserrat, Tothill, and Davey}}]{Mace1988}
\bibinfo{author}{\bibfnamefont{D.~A.~H.} \bibnamefont{Mace}},
  \bibinfo{author}{\bibfnamefont{D.~C.} \bibnamefont{Rogers}},
  \bibinfo{author}{\bibfnamefont{K.~J.} \bibnamefont{Monserrat}},
  \bibinfo{author}{\bibfnamefont{J.~N.} \bibnamefont{Tothill}},
  \bibnamefont{and} \bibinfo{author}{\bibfnamefont{S.~T.} \bibnamefont{Davey}},
  \bibinfo{journal}{Sernicond. Sci. Technol.} \textbf{\bibinfo{volume}{3}},
  \bibinfo{pages}{597} (\bibinfo{year}{1988}).

\bibitem[{\citenamefont{Vurgaftman et~al.}(2001)\citenamefont{Vurgaftman,
  Meyer, and Ram-Mohan}}]{Vurgaftman2001}
\bibinfo{author}{\bibfnamefont{I.}~\bibnamefont{Vurgaftman}},
  \bibinfo{author}{\bibfnamefont{J.~R.} \bibnamefont{Meyer}}, \bibnamefont{and}
  \bibinfo{author}{\bibfnamefont{L.~R.} \bibnamefont{Ram-Mohan}},
  \bibinfo{journal}{J. Appl. Phys.} \textbf{\bibinfo{volume}{89}},
  \bibinfo{pages}{5815} (\bibinfo{year}{2001}).

\bibitem[{\citenamefont{Arent et~al.}(1989)\citenamefont{Arent, Deneffe, Hoof,
  Boeck, and Borghs}}]{Arent1989}
\bibinfo{author}{\bibfnamefont{D.~J.} \bibnamefont{Arent}},
  \bibinfo{author}{\bibfnamefont{K.}~\bibnamefont{Deneffe}},
  \bibinfo{author}{\bibfnamefont{C.~V.} \bibnamefont{Hoof}},
  \bibinfo{author}{\bibfnamefont{J.~D.} \bibnamefont{Boeck}}, \bibnamefont{and}
  \bibinfo{author}{\bibfnamefont{G.}~\bibnamefont{Borghs}},
  \bibinfo{journal}{J. Appl. Phys.} \textbf{\bibinfo{volume}{66}},
  \bibinfo{pages}{1739} (\bibinfo{year}{1989}).

\bibitem[{\citenamefont{Chuang}(1995)}]{Chuang1995}
\bibinfo{author}{\bibfnamefont{S.~L.} \bibnamefont{Chuang}},
  \emph{\bibinfo{title}{Physics of Optoelectronic Devices}}
  (\bibinfo{publisher}{John Wiley \& Sons. Inc.}, \bibinfo{address}{New York},
  \bibinfo{year}{1995}).

\bibitem[{\citenamefont{Poggio et~al.}(2005)\citenamefont{Poggio, Myers, Stern,
  Gossard, and Awschalom}}]{Poggio2005}
\bibinfo{author}{\bibfnamefont{M.}~\bibnamefont{Poggio}},
  \bibinfo{author}{\bibfnamefont{R.~C.} \bibnamefont{Myers}},
  \bibinfo{author}{\bibfnamefont{N.~P.} \bibnamefont{Stern}},
  \bibinfo{author}{\bibfnamefont{A.~C.} \bibnamefont{Gossard}},
  \bibnamefont{and} \bibinfo{author}{\bibfnamefont{D.~D.}
  \bibnamefont{Awschalom}}, \bibinfo{journal}{Phys. Rev. B}
  \textbf{\bibinfo{volume}{72}}, \bibinfo{pages}{235313}
  (\bibinfo{year}{2005}).

\bibitem[{\citenamefont{Nazmula et~al.}(2003)\citenamefont{Nazmula, Sugaharaa,
  and Tanak}}]{Nazmula2003}
\bibinfo{author}{\bibfnamefont{A.~M.} \bibnamefont{Nazmula}},
  \bibinfo{author}{\bibfnamefont{S.}~\bibnamefont{Sugaharaa}},
  \bibnamefont{and} \bibinfo{author}{\bibfnamefont{M.}~\bibnamefont{Tanak}},
  \bibinfo{journal}{J. Cryst. Growth} \textbf{\bibinfo{volume}{251}},
  \bibinfo{pages}{303} (\bibinfo{year}{2003}).

\bibitem[{\citenamefont{Wurstbauer et~al.}(2009)\citenamefont{Wurstbauer, Soda,
  Jakiela, Schuh, Weiss, Zweck, and Wegscheider}}]{Wurstbauer2009}
\bibinfo{author}{\bibfnamefont{U.}~\bibnamefont{Wurstbauer}},
  \bibinfo{author}{\bibfnamefont{M.}~\bibnamefont{Soda}},
  \bibinfo{author}{\bibfnamefont{R.}~\bibnamefont{Jakiela}},
  \bibinfo{author}{\bibfnamefont{D.}~\bibnamefont{Schuh}},
  \bibinfo{author}{\bibfnamefont{D.}~\bibnamefont{Weiss}},
  \bibinfo{author}{\bibfnamefont{J.}~\bibnamefont{Zweck}}, \bibnamefont{and}
  \bibinfo{author}{\bibfnamefont{W.}~\bibnamefont{Wegscheider}},
  \bibinfo{journal}{J. Cryst. Growth} \textbf{\bibinfo{volume}{311}},
  \bibinfo{pages}{2160} (\bibinfo{year}{2009}).

\bibitem[{\citenamefont{Ando and Uemura}(1974)}]{Ando1974}
\bibinfo{author}{\bibfnamefont{T.}~\bibnamefont{Ando}} \bibnamefont{and}
  \bibinfo{author}{\bibfnamefont{Y.}~\bibnamefont{Uemura}},
  \bibinfo{journal}{J. Phys. Soc. Jpn.} \textbf{\bibinfo{volume}{36}},
  \bibinfo{pages}{959} (\bibinfo{year}{1974}).

\bibitem[{\citenamefont{Sampaio et~al.}(1997)\citenamefont{Sampaio, Freire, and
  Alves}}]{sampaio_alves_jap1997}
\bibinfo{author}{\bibfnamefont{J.~F.} \bibnamefont{Sampaio}},
  \bibinfo{author}{\bibfnamefont{S.~L.~S.} \bibnamefont{Freire}},
  \bibnamefont{and} \bibinfo{author}{\bibfnamefont{E.~S.} \bibnamefont{Alves}},
  \bibinfo{journal}{J. Appl. Phys.} \textbf{\bibinfo{volume}{81}},
  \bibinfo{pages}{530} (\bibinfo{year}{1997}).

\bibitem[{\citenamefont{Vosko et~al.}(1980)\citenamefont{Vosko, Wilk, and
  Nusair}}]{Vosko1980}
\bibinfo{author}{\bibfnamefont{H.~S.} \bibnamefont{Vosko}},
  \bibinfo{author}{\bibfnamefont{L.}~\bibnamefont{Wilk}}, \bibnamefont{and}
  \bibinfo{author}{\bibfnamefont{M.}~\bibnamefont{Nusair}},
  \bibinfo{journal}{Can. J. Phys.} \textbf{\bibinfo{volume}{58}},
  \bibinfo{pages}{1200} (\bibinfo{year}{1980}).

\bibitem[{\citenamefont{Perdew and Wang}(1992)}]{perdew_wang_prb1992}
\bibinfo{author}{\bibfnamefont{J.~P.} \bibnamefont{Perdew}} \bibnamefont{and}
  \bibinfo{author}{\bibfnamefont{Y.}~\bibnamefont{Wang}},
  \bibinfo{journal}{Phys. Rev. B} \textbf{\bibinfo{volume}{45}},
  \bibinfo{pages}{13244} (\bibinfo{year}{1992}).

\bibitem[{\citenamefont{Larson et~al.}(1988)\citenamefont{Larson, Hass,
  Ehrenreich, and Carlsson}}]{larson_carlson_prb1988}
\bibinfo{author}{\bibfnamefont{B.~E.} \bibnamefont{Larson}},
  \bibinfo{author}{\bibfnamefont{K.~C.} \bibnamefont{Hass}},
  \bibinfo{author}{\bibfnamefont{H.}~\bibnamefont{Ehrenreich}},
  \bibnamefont{and} \bibinfo{author}{\bibfnamefont{A.~E.}
  \bibnamefont{Carlsson}}, \bibinfo{journal}{Phys. Rev. B}
  \textbf{\bibinfo{volume}{37}}, \bibinfo{pages}{4137} (\bibinfo{year}{1988}).

\bibitem[{\citenamefont{Kacman}(2001)}]{Kacman2001}
\bibinfo{author}{\bibfnamefont{P.}~\bibnamefont{Kacman}},
  \bibinfo{journal}{Semicond. Sci. Technol.} \textbf{\bibinfo{volume}{16}},
  \bibinfo{pages}{R25} (\bibinfo{year}{2001}).

\bibitem[{\citenamefont{Dietl et~al.}(2001)\citenamefont{Dietl, Ohno, and
  Matsukura}}]{Dietl2001}
\bibinfo{author}{\bibfnamefont{T.}~\bibnamefont{Dietl}},
  \bibinfo{author}{\bibfnamefont{H.}~\bibnamefont{Ohno}}, \bibnamefont{and}
  \bibinfo{author}{\bibfnamefont{F.}~\bibnamefont{Matsukura}},
  \bibinfo{journal}{Phys. Rev. B} \textbf{\bibinfo{volume}{63}},
  \bibinfo{pages}{195205} (\bibinfo{year}{2001}).

\bibitem[{\citenamefont{Ashcroft and Mermin}(1976)}]{Ashcroft}
\bibinfo{author}{\bibfnamefont{N.~W.} \bibnamefont{Ashcroft}} \bibnamefont{and}
  \bibinfo{author}{\bibfnamefont{N.~D.} \bibnamefont{Mermin}},
  \emph{\bibinfo{title}{Solid State Physics}} (\bibinfo{publisher}{Saunders
  College, Philadelphia}, \bibinfo{year}{1976}).

\bibitem[{\citenamefont{Degani and Maialle}(2010)}]{Degani2010}
\bibinfo{author}{\bibfnamefont{M.~H.} \bibnamefont{Degani}} \bibnamefont{and}
  \bibinfo{author}{\bibfnamefont{M.~Z.} \bibnamefont{Maialle}},
  \bibinfo{journal}{J. Comput. Theor. Nanosci.} \textbf{\bibinfo{volume}{7}},
  \bibinfo{pages}{454} (\bibinfo{year}{2010}).

\bibitem[{\citenamefont{Bauer and Ando}(1986)}]{Bauer1986}
\bibinfo{author}{\bibfnamefont{G.~E.~W.} \bibnamefont{Bauer}} \bibnamefont{and}
  \bibinfo{author}{\bibfnamefont{T.}~\bibnamefont{Ando}}, \bibinfo{journal}{J.
  Phys. C: Solid State Phys.} \textbf{\bibinfo{volume}{19}},
  \bibinfo{pages}{1537} (\bibinfo{year}{1986}).

\bibitem[{\citenamefont{Bobbert et~al.}(1997)\citenamefont{Bobbert,
  Wieldraaijer, van~der Weide, Kemerink, Koenraad, and Wolter}}]{Bobbert1997}
\bibinfo{author}{\bibfnamefont{P.~A.} \bibnamefont{Bobbert}},
  \bibinfo{author}{\bibfnamefont{H.}~\bibnamefont{Wieldraaijer}},
  \bibinfo{author}{\bibfnamefont{R.}~\bibnamefont{van~der Weide}},
  \bibinfo{author}{\bibfnamefont{M.}~\bibnamefont{Kemerink}},
  \bibinfo{author}{\bibfnamefont{P.~M.} \bibnamefont{Koenraad}},
  \bibnamefont{and} \bibinfo{author}{\bibfnamefont{J.~H.}
  \bibnamefont{Wolter}}, \bibinfo{journal}{Phys. Rev. B}
  \textbf{\bibinfo{volume}{56}}, \bibinfo{pages}{3664} (\bibinfo{year}{1997}).

\bibitem[{\citenamefont{Wimbauer et~al.}(1994)\citenamefont{Wimbauer,
  Oettinger, Efros, Meyer, and Brugger}}]{wimbauer_brugger_prb1994}
\bibinfo{author}{\bibfnamefont{T.}~\bibnamefont{Wimbauer}},
  \bibinfo{author}{\bibfnamefont{K.}~\bibnamefont{Oettinger}},
  \bibinfo{author}{\bibfnamefont{A.~L.} \bibnamefont{Efros}},
  \bibinfo{author}{\bibfnamefont{B.~K.} \bibnamefont{Meyer}}, \bibnamefont{and}
  \bibinfo{author}{\bibfnamefont{H.}~\bibnamefont{Brugger}},
  \bibinfo{journal}{Phys. Rev. B} \textbf{\bibinfo{volume}{50}},
  \bibinfo{pages}{8889} (\bibinfo{year}{1994}).

\bibitem[{\citenamefont{Okabayashi et~al.}(1998)\citenamefont{Okabayashi,
  Kimura, Rader, Mizokawa, Fujimori, Hayashi, and
  Tanaka}}]{okabayashi_tanaka_prb1998}
\bibinfo{author}{\bibfnamefont{J.}~\bibnamefont{Okabayashi}},
  \bibinfo{author}{\bibfnamefont{A.}~\bibnamefont{Kimura}},
  \bibinfo{author}{\bibfnamefont{O.}~\bibnamefont{Rader}},
  \bibinfo{author}{\bibfnamefont{T.}~\bibnamefont{Mizokawa}},
  \bibinfo{author}{\bibfnamefont{A.}~\bibnamefont{Fujimori}},
  \bibinfo{author}{\bibfnamefont{T.}~\bibnamefont{Hayashi}}, \bibnamefont{and}
  \bibinfo{author}{\bibfnamefont{M.}~\bibnamefont{Tanaka}},
  \bibinfo{journal}{Phys. Rev. B} \textbf{\bibinfo{volume}{58}},
  \bibinfo{pages}{R4211} (\bibinfo{year}{1998}).

\bibitem[{\citenamefont{Kotlyar et~al.}(2001)\citenamefont{Kotlyar, Reinecke,
  Bayer, and Forchel}}]{kotlyar_forchel_prb2001}
\bibinfo{author}{\bibfnamefont{R.}~\bibnamefont{Kotlyar}},
  \bibinfo{author}{\bibfnamefont{T.~L.} \bibnamefont{Reinecke}},
  \bibinfo{author}{\bibfnamefont{M.}~\bibnamefont{Bayer}}, \bibnamefont{and}
  \bibinfo{author}{\bibfnamefont{A.}~\bibnamefont{Forchel}},
  \bibinfo{journal}{Phys. Rev. B} \textbf{\bibinfo{volume}{63}},
  \bibinfo{pages}{085310} (\bibinfo{year}{2001}).

\end{thebibliography}
\end{document}